\documentclass[12pt]{iopart}

\usepackage{collapsecite}
\usepackage{graphicx}
\usepackage{iopams}
\usepackage{color}

\begin{document}


%
%

%

%

%

\title[Stability and Dynamics of matter-wave vortices]{Stability and dynamics of matter-wave vortices in the
presence of collisional inhomogeneities and dissipative perturbations}

\author{
S.\ Middelkamp$^1$, 
P. G.\ Kevrekidis$^2$, 
D. J.\ Frantzeskakis$^3$,
R.\ Carretero-Gonz\'{a}lez$^{4,}$\footnote{On sabbatical leave from:
Nonlinear Dynamical System Group ({\texttt{URL}: http://nlds.sdsu.edu}),
Computational Science Research Center,
and Department of Mathematics and Statistics,
San Diego State University, San Diego, California 92182-7720, USA},
and P.\ Schmelcher$^{1}$}
%
\address{$^1$ Zentrum f\"ur Optische Quantentechnologien, Universit\"at
Hamburg, Luruper Chaussee 149, 22761 Hamburg, Germany}
\address{$^2$ Department of Mathematics and Statistics, University of Massachusetts,
Amherst MA 01003-4515, USA}
\address{$^3$ Department of Physics, University of Athens, Panepistimiopolis,
Zografos, Athens 157 84, Greece}
%
%
\address{$^4$Nonlinear Physics Group,
Departamento de F\'{\i}sica Aplicada I,
Universidad de Sevilla,
Avda.~Reina Mercedes s/n., 41012 Sevilla, Spain}
%
%
\ead{stephan.middelkamp@physnet.uni-hamburg.de}

\begin{abstract}
In this work, the spectral properties of a singly-charged vortex in a
Bose-Einstein condensate confined in a highly anisotropic (disk-shaped)
harmonic trap are investigated. Special emphasis is given on the analysis
of the so-called anomalous (negative energy) mode of the Bogoliubov
spectrum. We use analytical and numerical techniques to illustrate
the connection of the anomalous mode to the precession dynamics of
the vortex in the trap. Effects due to inhomogeneous interatomic
interactions and dissipative perturbations motivated
  by finite temperature 
considerations are explored. We find that
both of these effects may give rise to oscillatory instabilities of the
vortex, which are suitably diagnosed through the perturbation-induced
evolution of the anomalous mode, and being monitored by direct
numerical simulations.
\end{abstract}

\pacs{03.75.Lm, 67.90.+z, 34.50.Cx}
\submitto{\JPB}
\maketitle


\section{Introduction}

Matter-wave vortices represent fundamental nonlinear macroscopic
excitations of Bose-Einstein condensates (BECs);
see e.g.~the relevant reviews 
\cite{stringari,pethick,emergent,fetter,rev_fetter,PGK:MPLB:04,review}.
These structures are characterized by their nonzero topological charge $S$,
the phase dislocation
and jump by $2 \pi S$ induced by the vorticity and the concomitant
vanishing of the BEC density at the vortex core.
Experimental observation of matter-wave vortices was first reported in
Ref.~\cite{Matthews99},
using a phase-imprinting method between two hyperfine spin states of a $^{87}$Rb BEC \cite{Williams99}.
Other techniques for the generation of vortices have also been studied theoretically and implemented in experiments. 
In particular,
stirring the BEC \cite{Madison00} above a certain critical angular speed \cite{Recati01,Sinha01,Madison01}
is an extremely efficient method for producing
a few vortices \cite{Madison01} or
vortex lattices \cite{Raman}.
Other techniques include
the supercritical dragging of an obstacle through the BEC \cite{frisch92,Jackson98,kett99},
as well as the nonlinear interference of condensate fragments
\cite{BPAPRL,Nate1,Nate2,Nate3}.
In the above studies, vortices were singly-charged i.e., with a topological charge $S=1$;
higher-charged vortices with $S>1$
may also be created experimentally \cite{Leanhardt,S2Ket} and could, in principle, be stable under appropriate
conditions \cite{Pu99,S2Mottonen}. 
Considerable effort has been dedicated to the investigation of the
stability of such higher charge structures \cite{pra_s2a,pra_s2b,pra_s2c,pra_s2d}.
Nevertheless, such higher-charged vortices
are typically far less robust than the fundamental $S=1$
vortex that is of interest here.

In this work we systematically study singly-charged vortices in a two-dimensional
(2D) ---so-called disk-shaped--- BEC
from a spectral (i.e., Bogoliubov-de Gennes) point of view.
In particular, first we focus on the
so-called anomalous mode of the Bogoliubov theory, characterized by negative energy
\cite{feder} or negative Krein-sign \cite{bjorn}, and elucidate its
connection with the precessional motion of the vortex,
if displaced from its equilibrium position i.e., the trap center.
Next, we will study how this mode is affected by the presence of different
kinds of perturbations.
The perturbations we consider here arise from inhomogeneous interatomic
interactions, so-called collisional inhomogeneities, and finite-temperature
induced dissipation.

Interatomic interactions, characterized by the $s$-wave scattering length, become
spatially (or temporally) varying
by employing either magnetic \cite{Koehler,feshbachNa} or optical Feshbach
resonances \cite{ofr,rempe1,rempe2} in a very broad range. This remarkable flexibility
on the manipulation of the effective mean-field nonlinearity of BECs, has
inspired a significant number of experimental and theoretical studies.
Herein, we will focus on the more recently proposed
``collisionally inhomogeneous'' BECs, characterized by a
spatially-dependent scattering length.
In such settings, many interesting phenomena have been predicted, including
adiabatic compression of matter-waves \cite{our1,fka},
Bloch oscillations of solitons \cite{our1},
soliton emission and atom lasers \cite{vpg12},
enhancement of transmittivity of matter-waves through barriers \cite{our2,fka2},
dynamical trapping of solitons \cite{our2},
stable condensates exhibiting both attractive and repulsive interatomic
interactions \cite{chin} and the
delocalization transition of matter waves \cite{LocDeloc}.
Here we will examine how harmonic spatial variations of the scattering length,
inducing a sort of a nonlinear optical lattice in the system,
affect the stability and ensuing dynamics of the vortex.
Interestingly, we find that the anomalous mode of the vortex  (located at the origin) is differently
affected by cosinusoidally (the vortex is located at a
  maximum of the nonlinearity) and sinusoidally (the
  vortex is located at a local minimum of the nonlinearity) varying
nonlinearities i.e., the phase of the nonlinearity's spatial variation
at the vortex location plays a
crucial role in the ensuing stability properties.
This turns out to be the most critical element of influence
within this setting. In the former case the vortex is stable,
while in the latter the vortex is subject to an oscillatory instability, emerging by
the collision of the anomalous mode with another eigenmode of the system.

We also consider in our study the effect of
  dissipative perturbations on the vortex dynamics motivated by
  considerations of the coherent structure's interaction with the thermal cloud. Here we will adopt a simple phenomenological
model relying on the inclusion of a phenomenological damping in the mean-field model, first
introduced by Pitaevskii \cite{lp} and subsequently used in various works to describe,
e.g., decoherence \cite{graham} and growth \cite{gard2} of BECs,
damping of collective excitations of BECs \cite{choi},
vortex lattice growth \cite{ueda,gard}, vortex dynamics \cite{mad} (see also \cite{Nate1,Nate2}),
and decay of dark solitons \cite{nppprl}.
Importantly, inclusion of such a phenomenological damping in the Gross-Pitaevskii equation (GPE)
can be justified from a microscopic perspective (see, e.g., the recent review \cite{npprev}).
Herein, we will show how such a finite-temperature motivated dissipation
affects the statics and dynamics of the vortex, by leading its anomalous
mode to become immediately unstable. Despite the
  relatively simple and phenomenological nature of the model, we will
  see that its results will bear significant similarities to the
  phenomenology of more complex dynamical models of the relevant
  interactions, allowing us to understand qualitatively the origin of the observed
  dynamical features.
We will also present some interesting twists that may arise when the combined effect
of thermal dissipation and spatially-dependent interatomic interactions is considered.

The paper is structured as follows. In Section \ref{secII}, we present our analytical considerations
in connection to the spectrum of a $S=1$ vortex and its precession frequency in the trap.
In Section \ref{secIII}, we examine numerically the validity of the analytical
predictions, but also how these are modified in the presence of additional perturbations
such as the spatially dependent nonlinearity, or the finite-temperature
induced dissipative perturbation. This is done both through
a systematic analysis of the Bogoliubov-de Gennes equations,
as well as through the direct numerical simulations of the pertinent GPE models.
Finally, in Section \ref{secIV},
we summarize our results and discuss directions for future studies.

\section{Analytical Results}
\label{secII}
We first consider the simplest case in our study, namely a matter-wave vortex in a 2D BEC confined in
a harmonic trap. It is well-known that in this setting the singly charged vortex will have precisely one
anomalous mode \cite{feder}; this mode, characterized by a negative energy, is also
known as mode of negative Krein sign (or signature) in the
mathematical literature \cite{bjorn}; see also below for an explicit
mathematical definition.
In Ref.~\cite{feder} (see also the review \cite{fetter}), it was argued that the single negative energy mode
with $S=1$ (which is of interest here) is responsible for the precessional motion of the vortex in the trap
(in addition to being relevant for other processes such as vortex nucleation).

One of the key purposes in our study is to consider the precession in the setting described above.
The three-dimensional (3D) analogue of this setting has been considered and studied analytically
by means of the matched asymptotics technique in Ref.~\cite{fetter1},
while the 2D case has been studied by means of a variational approach in Ref.~\cite{kim} (see
more details below). Here, employing the matched asymptotics method, we derive an
expression for the precession frequency in the 2D case, and provide a detailed comparison of
this result with numerics pertaining to the study of the anomalous mode.

The model under consideration is the $(2+1)$-dimensional GPE \cite{review},
\begin{eqnarray}
i \hbar \partial_{t^\prime} u = - \frac{\hbar^2}{2m} \Delta^\prime u + V(r^\prime) u + g_{2D} |u|^2 u-\mu^\prime u.
\label{veq1}
\end{eqnarray}
Here, $u(x^\prime,y^\prime,t^\prime)$ is the macroscopic wave function of the disk-shaped BEC, $\Delta^\prime$
is the 2D Laplacian, $r^\prime \equiv \sqrt{x^{\prime2}+y^{\prime2}}$ is the radial variable,
$V(r^{\prime})=\frac{1}{2} \omega_r^2 r^{\prime2}$ is the harmonic trapping potential in the in-plane direction, $\mu^\prime$ the chemical potential and
$g_{2D}=g_{3D}/2\pi a_z=2\sqrt{2\pi} a a_z \hbar \omega_z$ is the effectively 2D nonlinear coefficient where $a$ is the 
scattering length and $a_z$, $\omega_z$ are the transverse
(strongly confining)
harmonic oscillator length and trapping frequency, respectively. 
Measuring length in units of $a_z$ and frequencies in units of
$\omega_z$ 
Eq.~(\ref{veq1}) can be expressed in the following dimensionless form,
\begin{eqnarray}
i \partial_t u = - \frac{1}{2} \Delta u + V(r) u + g |u|^2 u-\mu u.
\label{veq1b}
\end{eqnarray}
Here, $\Delta$ is the 2D Laplacian of the rescaled variables, $r$ is the rescaled radial variable, $V(r)=\frac{1}{2} \Omega^2 r^2$ is the harmonic trapping potential with
$\Omega$ being measured in units of $\omega_z$,
$g$ is the normalized strength of the interatomic interactions (which we set to $g=1$
for the analytical considerations of this section),
and $\mu$ is the chemical potential measured in units of $\hbar\omega_z$.
In order to study the effect of the potential on the vortex, we will follow the lines of
Ref.~\cite{pis1} (see also Ref.~\cite{pis2} for similar work in the context of optics)
and use a matched asymptotics approach between an inner and
an outer perturbative solution leading to the following equations of motion
(for a more detailed derivation see the appendix)
for the vortex center $(x_v,y_v)$:
\begin{eqnarray}
\dot{x}_v&=&\phantom{-}\frac{\Omega^2}{2 \mu} \log \left(A \frac{\mu}{\Omega}\right) y_v,
\label{veq8}
\\
\dot{y}_v&=&-\frac{\Omega^2}{2 \mu} \log \left(A \frac{\mu}{\Omega}\right) x_v,
\label{veq9}
\end{eqnarray}
where $A$ is an appropriate numerical factor (detailed comparison
with numerics yields very good agreement in the Thomas-Fermi
regime e.g.~for $A \approx 8.88\approx 2 \sqrt{2}\pi$, see
below). These results
suggest a precession of the vortex
in the harmonic trap with a frequency
\begin{eqnarray}
\omega_{\rm an}= \frac{\Omega^2}{2 \mu} \log(A \frac{\mu}{\Omega}), 
\label{veq10}
\end{eqnarray}
which, as suggested by the subscript (``an'' stands for anomalous),
should coincide with the eigenfrequency of the anomalous
mode of the Bogoliubov spectrum.
The anomalous mode eigenfrequency can readily be
  obtained through a standard
Bogoliubov-de Gennes (BdG) analysis. This analysis involves
the derivation of the BdG equations, which stem from a linearization
of the GPE (\ref{veq1}) around the vortex solution $u_0$ by using the ansatz
\begin{eqnarray}
u =u_0(x,y) 
+ \left[a(x,y) e^{i \omega t}
+ b^{\star}(x,y) e^{-i \omega^{\star} t} \right].
\label{veq11}
\end{eqnarray}
The subsequent solution of the ensuing BdG
eigenproblem yields the eigenfunctions $\{a(x,y),b(x,y)\}$ and
eigenfrequencies $\omega$.

Note that due to the
Hamiltonian nature of the system, if $\omega$ is an eigenfrequency
of the Bogoliubov spectrum, so are $-\omega$, $\omega^{\ast}$ and
$-\omega^{\ast}$. Notice that a linearly stable configuration is
tantamount to ${\rm Im}(\omega) =0$, i.e., all eigenfrequencies being real.

An important quantity resulting from the BdG analysis is the amount of energy carried by the normal mode with eigenfrequency $\omega$, namely
%
\begin{equation}
E=\int{dxdy(|a|^2-|b|^2)}
\omega.
\label{energy}
\end{equation}
The sign of this quantity, known as {\it Krein sign} \cite{MacKay},
is a topological property of each eigenmode. 
For one of the eigenvalues of each double pair this sign is negative. The corresponding mode is called {\it negative energy mode} (in the physical literature) \cite{skryabin} or mode with {\it negative Krein signature} (in the mathematical literature) \cite{MacKay}.
Practically, this means if it becomes resonant with a mode with positive Krein signature then, in most cases, 
there appear complex frequencies in the excitation spectrum, i.e., a dynamical instability occurs \cite{MacKay}.
The eigenvalues with negative Krein signature are actually associated
with the anomalous modes \cite{stringari} appearing
in the BdG spectrum.

In order to compare our results
to the ones obtained in earlier works, we should mention that
a similar setup was investigated in Ref.~\cite{McGee} for finite displacements of the vortex from the center of the trap and in Ref.~\cite{kim}
(by means of a variational approximation). In the latter one
the frequency of the anomalous mode was derived with
a similar functional form. However, in Ref.~\cite{kim},
the case of a BEC unbounded in the axial direction ($\omega_z=0$)
was considered and, as a result,
the constant was found to take a different value, $A=2$.
It is also worth noting that the connection between
quantum fluctuations and anomalous modes of matter-wave vortices under
Magnus forces was considered in Ref.~\cite{DziarmagaMeisner05}


It is important, at this stage, to make a few comments
regarding the nature of the Bogoliubov spectrum
resulting from the linearization around the vortex.
The system at hand, namely the disk-shaped condensate carrying
the vortex, is {\it not} in the ground state (a similar situation
occurs in the 1D analogue of the system, namely a quasi-1D BEC
carrying a dark soliton). The existence of the anomalous mode,
characterized by negative energy, indicates that the vortex
(and the dark soliton in the 1D case) is thermodynamically
unstable and, in the presence of dissipation, the system is
driven towards a lower energy configuration, namely the ground state.
Also, the eigenfrequency of the anomalous mode of the vortex
(similarly to the case of the dark soliton \cite{peli1}) bifurcates
from its value $\omega=\Omega$ (the linear oscillation with
the trap frequency) in the linear limit
---where the vortex is represented by the linear superposition
$|1,0 \rangle + i |0,1\rangle$, where $|m,n\rangle$ denotes
the $m$-th linear eigenstate of the quantum harmonic oscillator
along the $x$-direction and $n$-th one along the $y$-direction
(see discussion in Section \ref{secIIIA} and Fig.~\ref{fig1}). Generally,
the anomalous mode (in both 1D and 2D cases) is the {\it lowest}
excitation frequency of the system and the only one below
the trap frequency, which is associated with the
doubly degenerate (in the two-dimensional case) Kohn mode corresponding to
the dipolar motion of the condensate \cite{giorgini}.
However, in the case of the vortex
(and contrary to what is the case for the dark soliton),
there is one more frequency, which may
be smaller than $\omega_{\rm an}$, at least for small chemical potentials.
However the corresponding mode grows monotonically away from
the origin and thus becomes larger for increasing chemical potential than the anomalous mode and the Kohn mode.
This frequency was described through a small parameter expansion in Section
V.B of Ref.~\cite{todd2}. The relevant eigenfrequency is given
(in our units) by the following expression,
\begin{eqnarray}
\omega \approx \mu-2 \Omega,
\label{veq12}
\end{eqnarray}
which becomes increasingly more accurate as $\mu \rightarrow 2
\Omega$. This is in contrast to the case for
the expression of the precession frequency Eq.(\ref{veq10}),
which should be increasingly more accurate in the
Thomas-Fermi (TF) limit, corresponding to large $\mu$.

We now turn to numerical investigations in order to examine
the validity of our results in the case of the parabolic trap for constant
nonlinearity strength, as well as to generalize them to settings
which are less straightforward to consider by analytical means.
The results will be partitioned in two subsections: firstly, we
will provide bifurcation results from the BdG analysis, and
subsequently, we will also test the BdG predictions against
full numerical integration of Eq.~(\ref{veq1}).

\section{Numerical Results}
\label{secIII}

\subsection{Bogoliubov-de Gennes Analysis}
\label{secIIIA}

\begin{figure}[tbp]
\begin{center}
\includegraphics[width=8cm,angle=0,clip]{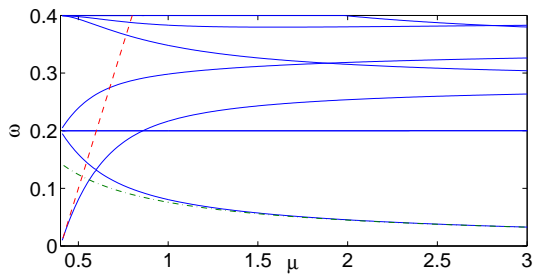}
\caption{(Color online) The eigenfrequency $\omega$ of the Bogoliubov spectrum
for a harmonically confined singly-charged vortex 
as a function of the chemical potential $\mu$ (for a trap strength $\Omega=0.2$).
Theoretical predictions are given by $\omega=(\mu-2 \Omega)$
[(red) dashed line] for the mode monotonically increasing from zero and
$\omega = (\Omega^2/(2 \mu)) \log(A \mu/\Omega)$ [(green) dash-dotted line];
the constant $A$ was chosen to be $2 \sqrt{2}\pi \approx 8.886$.}
\label{fig1}
\end{center}
\end{figure}

We start with the case of a singly-charged vortex in a
harmonically confined BEC
with homogeneous interatomic interactions (i.e., $g=1$).
In Fig.~\ref{fig1} we show the numerically obtained eigenfrequency $\omega$
of the Bogoliubov spectrum as (blue) solid lines as a function of the chemical
potential $\mu$ and the analytical predictions of Eq.~(\ref{veq10}) (green) dashed-dotted line and  Eq.~(\ref{veq12}) (red) dashed line. The numerically obtained frequencies are real denoting that the system is dynamically stable. We observe that
in accordance to the analytical predictions, the lowest modes are
(i) the one monotonically increasing away from zero and (ii) the anomalous
mode, connected to the vortex precession (see previous section), bifurcating from the constant dipolar mode in the liner limit. For the
monotonically increasing mode, we notice that the
non-radial nature of the solutions at hand (due to their
phase profile) leads to the absence of additional symmetries
of the eigenvalue problem away from the linear limit.
The only symmetry generally present is that of the phase or gauge
invariance, associated with the conservation of the number of particles.
This sustains a pair of linearization eigenfrequencies
at $\omega=0$, but as discussed in Ref.~\cite{todd2}, at the linear
limit the dimension of the corresponding kernel is $4$, hence
an eigenfrequency pair should depart from the origin
(at least for small $\mu$) according to Eq.~(\ref{veq12}).
As observed in Fig.~\ref{fig1}, this prediction
is in good agreement with the numerical results.
Naturally, deviations are observed
for larger chemical potentials. On the other hand, as
concerns
the precession frequency, we notice its monotonically decreasing
dependence on $\mu$ for given $\Omega$ (the latter, was set to
$\Omega=0.2$ in Fig.~\ref{fig1}),
its bifurcation from the Kohn mode eigenfrequency limit
and its excellent agreement with the theoretical prediction
in the TF limit
(for all $\mu > 1$). We note in passing that in this two-dimensional
case, a pair of Kohn modes can be seen to be preserved at
$\omega=\Omega=0.2$, being associated with the dipole oscillations
of the condensate along the two spatial directions, while the fourth
mode at $0.2$ in the linear limit results in a monotonically growing
eigenfrequency, as $\mu$ is increased.

\begin{figure}[tbp]
\begin{center}
\includegraphics[width=8.0cm,height=6cm,angle=0,clip]{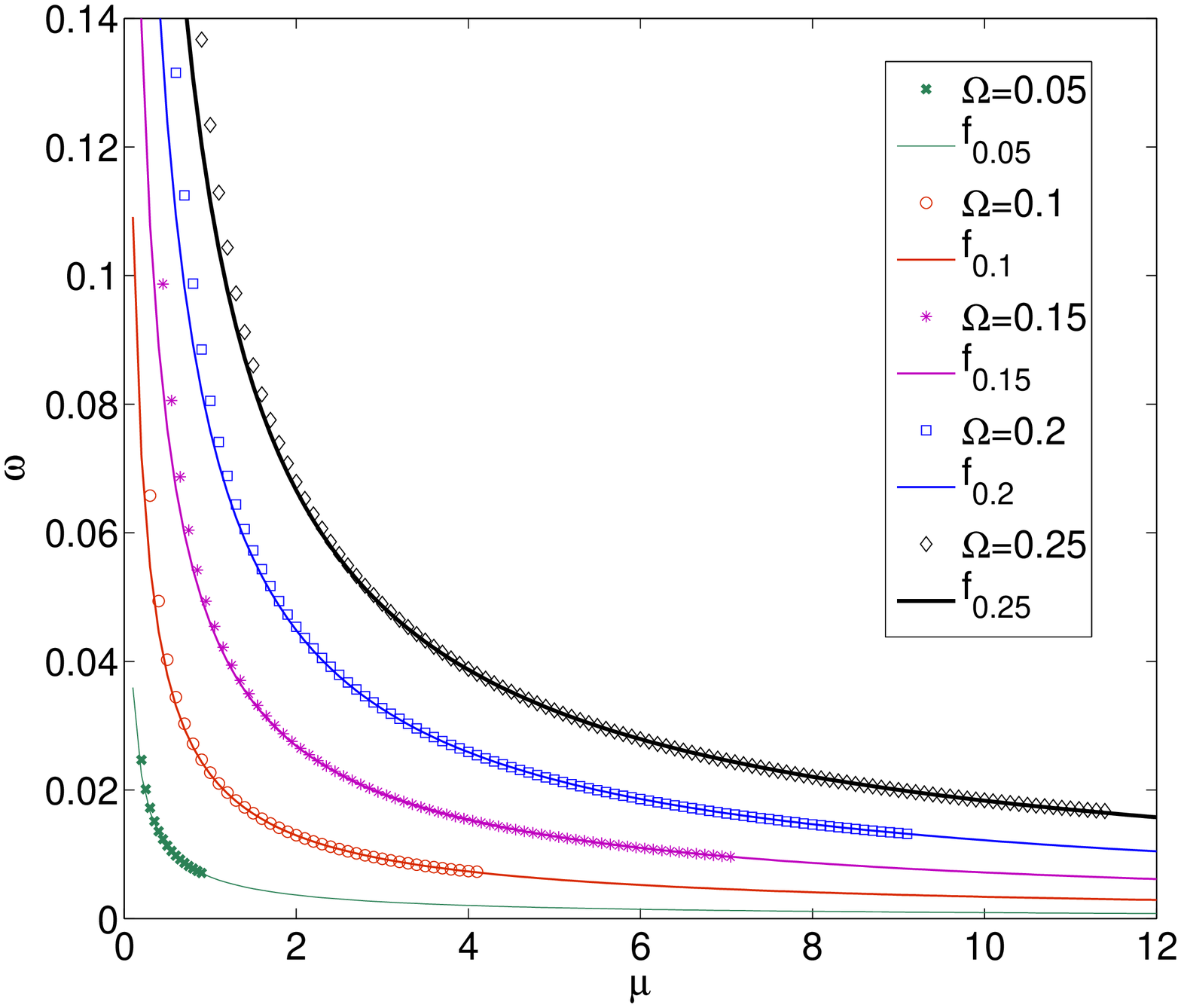}
\includegraphics[width=8.0cm,height=6cm,angle=0,clip]{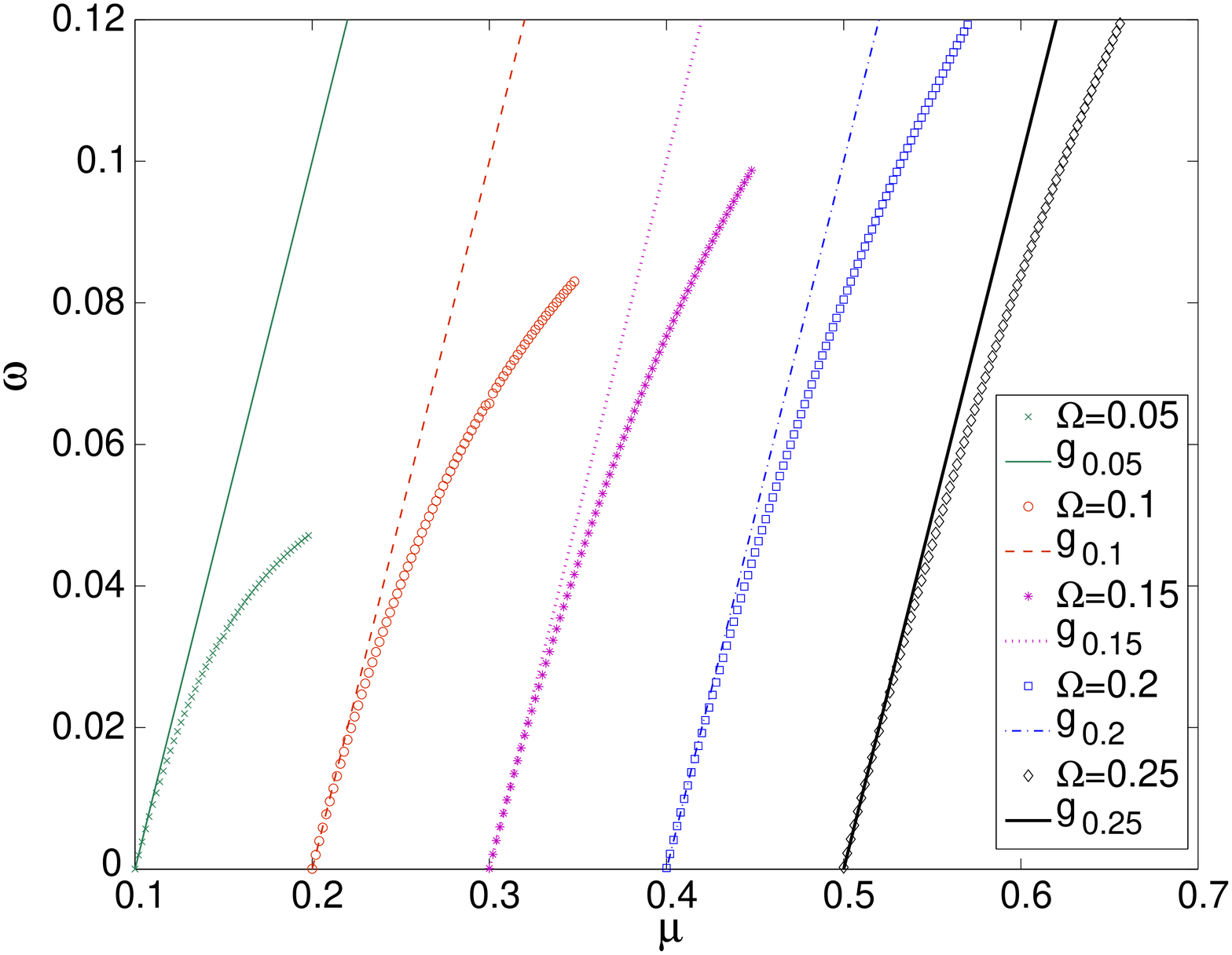}
\caption{(Color online) Top: dependence of the anomalous mode
eigenfrequency on the chemical potential for different trap
frequencies $\Omega$. The data points are interpolated by the functions
$f_\Omega(\mu)=(\Omega^2/(2 \mu)) \log(2 \sqrt{2}\pi \mu/\Omega)$.
Bottom: similar to top panel, but for
the mode bifurcating from zero, as compared to the theoretical
prediction $g_\Omega(\mu)=\mu-2 \Omega$.}
\label{fig1a}
\end{center}
\end{figure}

We have also tested the validity of the analytical predictions concerning
the two lowest eigenfrequencies
for different values of $\Omega$ and as a function of
$\mu$
(see Fig.~\ref{fig1a}). Once again,
a very good agreement of the two asymptotic theoretical descriptions
in their respective limits is found.

We now consider an interesting modification to this picture,
arising from a consideration of inhomogeneous interatomic
interactions, described by
a spatially-dependent scattering length $a(x,y)$
(see, e.g., the recent special
volume \cite{victor}). Here, we will consider
the effect of a periodic variation of the nonlinearity
strength, $g \equiv g(x,y)$ (i.e., a sort of a nonlinear optical
lattice) on the spectrum of a vortex. We will also
draw parallels with
similar spectral implications in the setting of
a linear periodic potential analyzed in Ref.~\cite{law}.

\begin{figure}[tbp]
\begin{center}
\includegraphics[width=8cm,angle=0,clip]{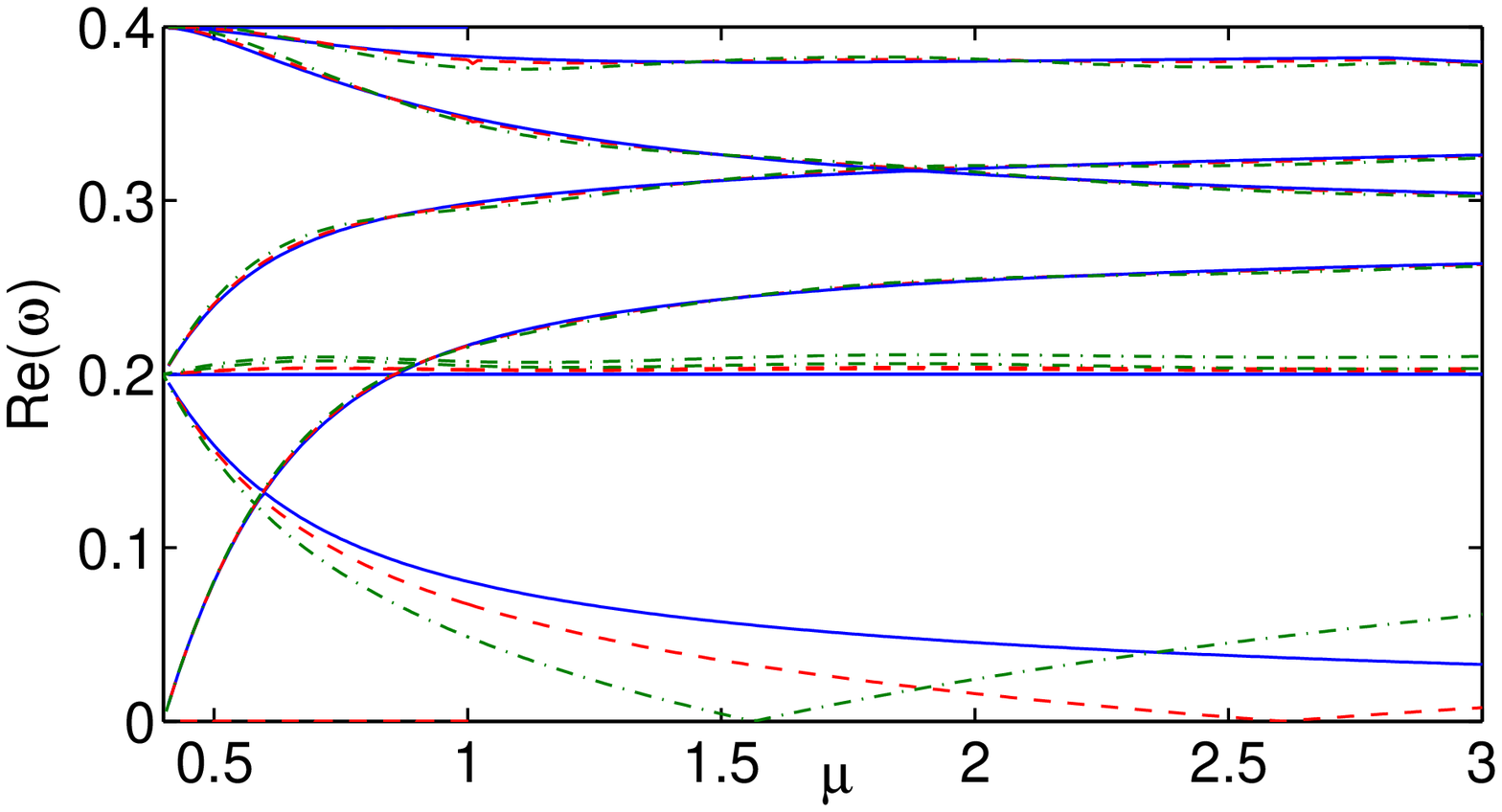}
\\
\includegraphics[width=8cm,angle=0,clip]{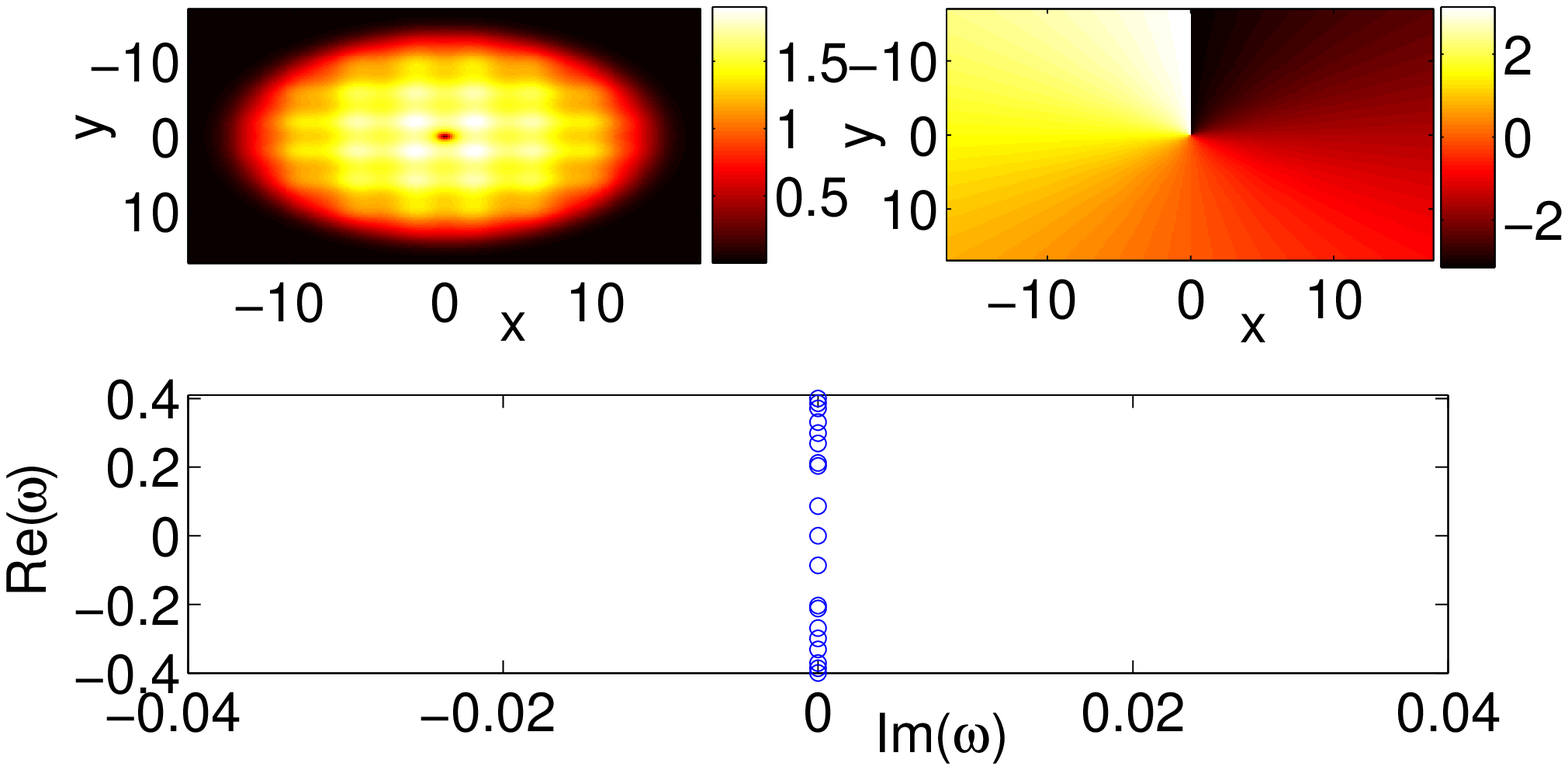}
\caption{(Color online) The case of a cosinusoidal variation of the
nonlinearity strength for a trap strength $\Omega=0.2$. The top panel is similar to Fig. \ref{fig1} showing 
the eigenfrequency $\omega$ of the Bogoliubov spectrum
as a function of the chemical potential $\mu$.
Comparison of $g(x,y)=1+ s \left(\cos^2(\pi x/4)
+ \cos^2(\pi y/4)\right)$, with $s=0.1$ [(red) dashed line] and
$s=0.3$ [(green) dash-dotted line], with the case of $s=0$ [(blue) solid
line). The middle panels show contour plots of the density (left)
and phase (right) of the wave function, while the bottom
panel shows the respective Bogoliubov excitation spectrum
(real vs. imaginary part of the eigenfrequency $\omega$,
where instability would correspond to the existence of
eigenfrequencies with Im$(\omega)\neq0$).
The chemical potential is $\mu=4$ (approaching
the Thomas-Fermi limit).
}
\label{fig2}
\end{center}
\end{figure}

In Fig.~\ref{fig2}, we
study the case of a
{\it cosinusoidal} variation of the nonlinearity strength,
namely, $g(x,y)=1+ s \left(\cos^2(\pi x/4) + \cos^2(\pi y/4)\right)$,
monitoring the vortex spectrum as a function of the chemical potential, where
$s$ is the strength of the oscillation.
In the same figure, the typical form of
the density and phase of the wave function (the former
showcasing spatial variation dictated by the corresponding
variation of the scattering length, and the latter demonstrating
the vortex structure of the configuration), as well as the
Bogoliubov excitation spectrum, are also illustrated.
We notice that while most of the relevant eigenfrequencies are only
very weakly affected by the spatially-dependent nonlinearity,
the one which is most {\it dramatically} affected is
that of the anomalous mode. The comparison of the $s=0$
case of Fig.~\ref{fig1} (blue solid lines) with the red dashed line
of $s=0.1$ and the green dash-dotted of $s=0.3$ illustrates
that the latter two not only approach zero, but rather cross
it at a finite value of $\mu$. For $s=0.1$, the anomalous mode
hits the origin of the spectral plane at $\mu=2.61$, while
for $s=0.3$ at $\mu=1.56$. However, it is
perhaps even more remarkable that this collision does not
produce an instability through an imaginary
eigenfrequency (real eigenvalue) pair, but rather maintains the stability
of the configuration (the eigenfrequencies appear to go through each
other). Generally, it can be seen that the trend of increasing
the oscillation strength in the cosinusoidal case leads to a more rapid decrease
of the anomalous mode eigenfrequency with $\mu$ and an ``earlier''
collision (i.e., occurring for smaller $\mu$) with the spectral plane origin.
The present study focuses on a periodicity
(wavelength) of the nonlinearity that is larger than the size (core) of the 
vortex. All throughout this long wavelength regime 
the spectral results are qualitatively 
the same. The case pertaining to wavelengths of the nonlinearity comparable 
or smaller than the size of the vortex falls outside of the scope
of the current manuscript and will be studied further in a future work.
Nonetheless, it can be anticipated that for small enough wavelengths
compared to the core of the vortex, the spatial modulation of the
nonlinearity will effectively, through spatial homogenization, 
act as a constant nonlinearity (possibly shifted from its
original $g=1$ value) in a manner akin to the effects of (linear) periodic
potentials generated by optical lattices acting on
harmonically trapped dark solitons \cite{dsol_A}.

\begin{figure}[tbp]
\begin{center}
\includegraphics[width=8cm,angle=0,clip]{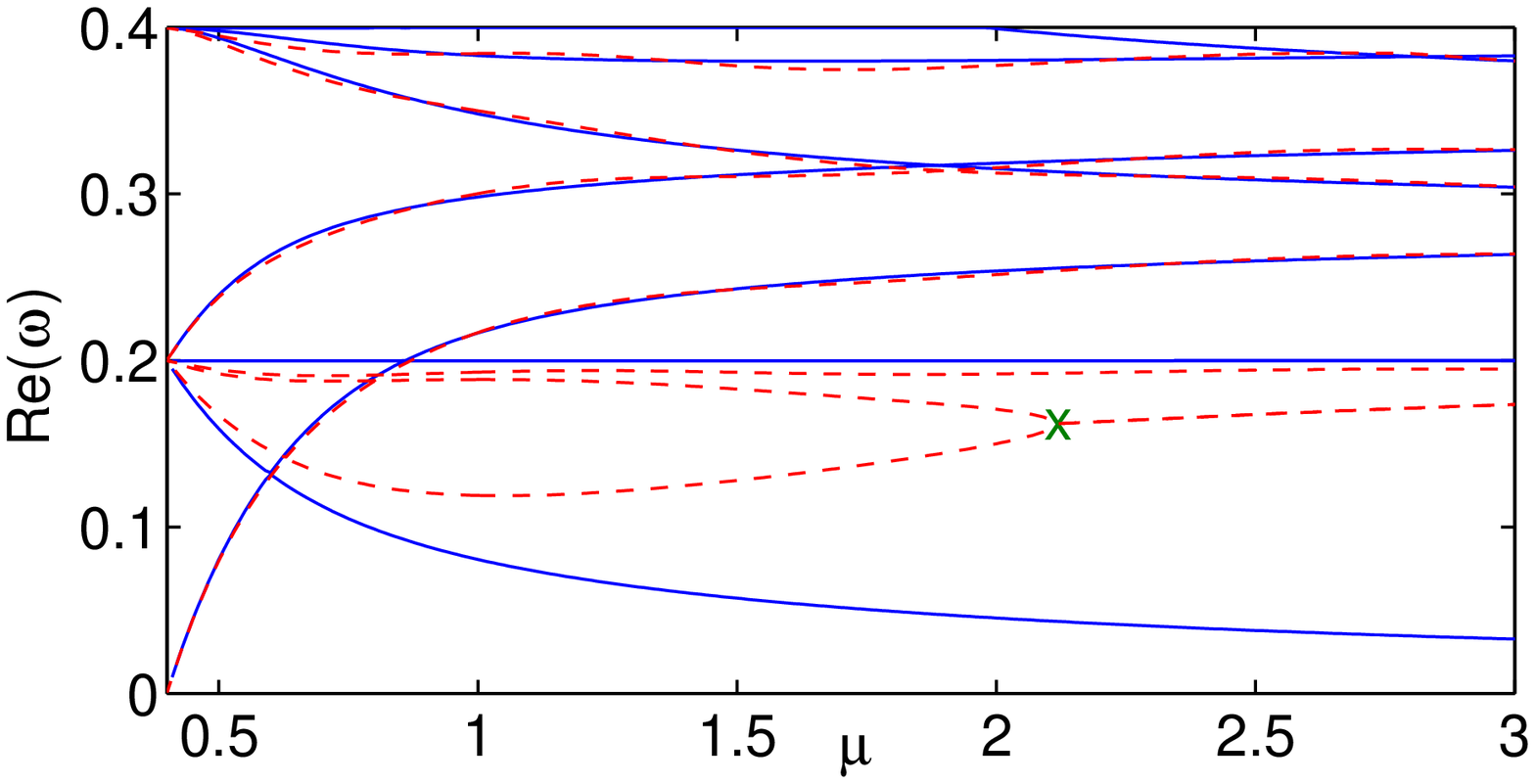}
\\
\includegraphics[width=8cm,angle=0,clip]{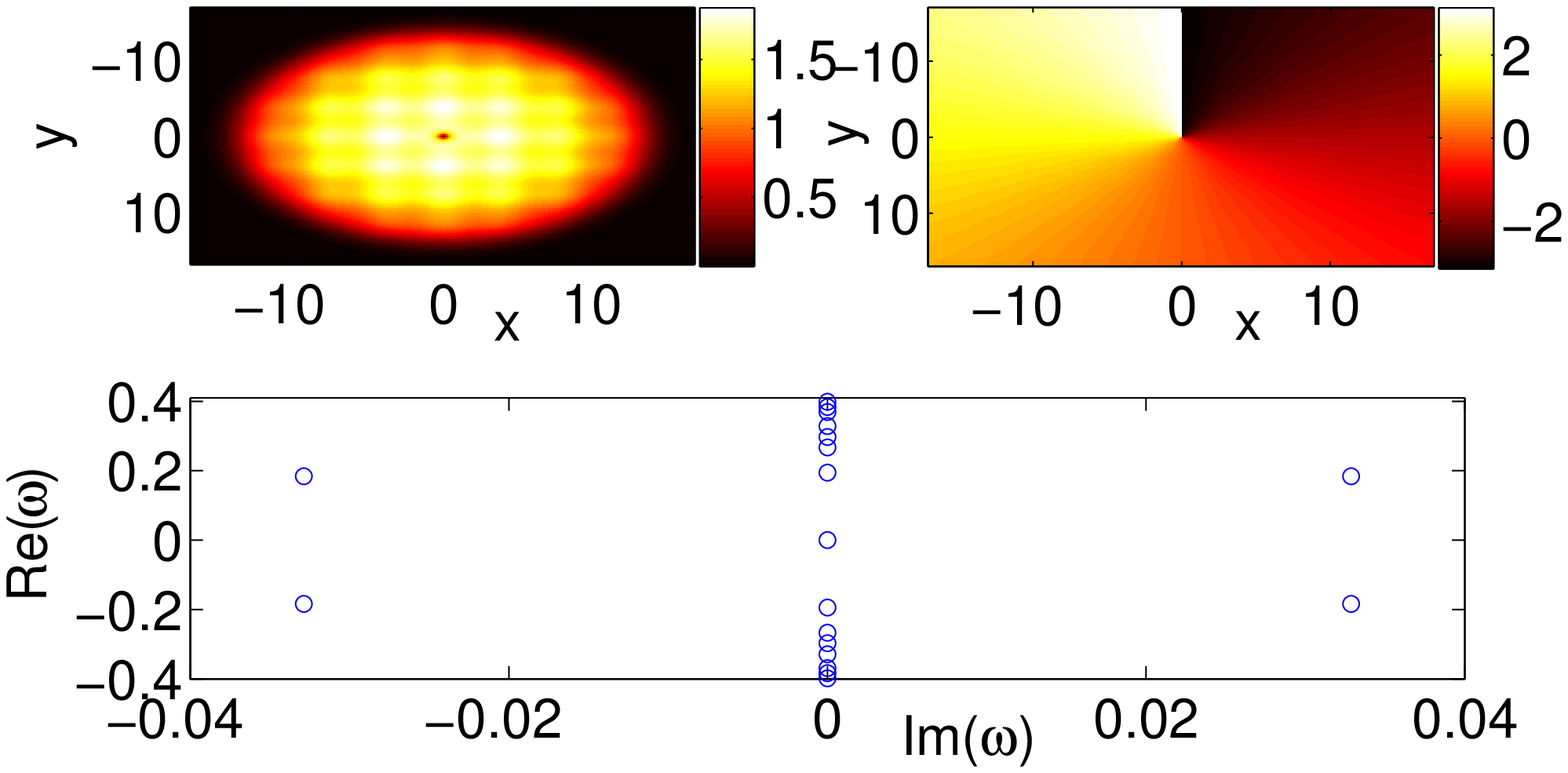}
\\[1.0ex]
\includegraphics[width=8cm,angle=0,clip]{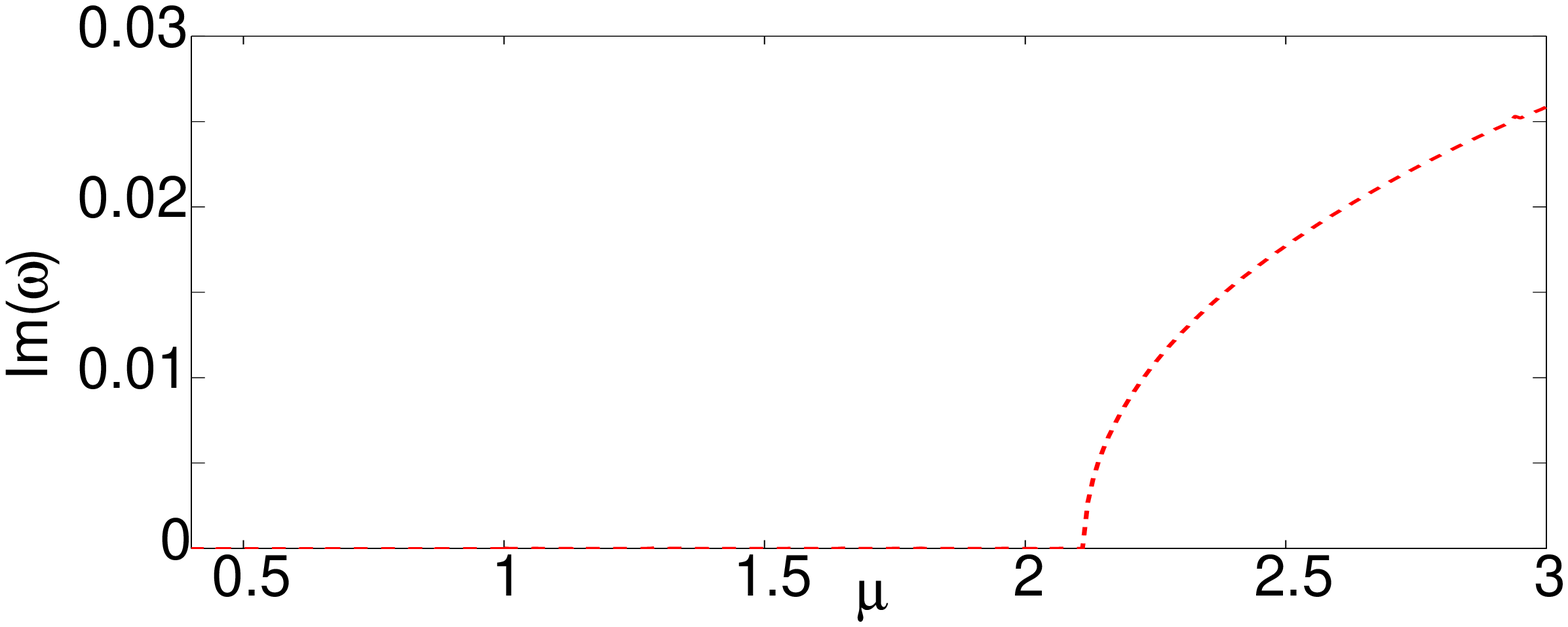}
\caption{
(Color online)
Similar to Fig.~\ref{fig2}, but for the case of a sinusoidal variation of the
nonlinearity strength. The first three rows of panels are analogous to the panels of
Fig.~\ref{fig2}. The case of $g(x,y)=1+ s \left(\sin^2(\pi x/4)
+ \sin^2(\pi y/4)\right)$, for $s=0.3$ [(red) dashed line] is compared
to that of $s=0$ [(blue) solid line]. The bottom panel shows the imaginary
part of the complex eigenfrequency;
the oscillatory instability arises for $\mu > 2.11$.}
\label{fig3}
\end{center}
\end{figure}

It is now interesting to turn to the case of the {\it sinusoidal}
modulation of the nonlinearity strength, namely
 $g(x,y)=1+ s \left(\sin^2(\pi x/4)
+ \sin^2(\pi y/4)\right)$. In this case, as observed in Fig.~\ref{fig3},
the fundamental difference is that the anomalous
mode is larger than that of the
homogeneous interactions case ($g=1$). More importantly
perhaps, its dependence can also be non-monotonic, resulting in the
increase of the corresponding eigenfrequency for a chemical
potential $\mu \gtrsim 1$.
Consequently, this raises
the possibility of collision of the relevant eigenmode with
other modes bifurcating from $\omega=\Omega$ for sufficiently
large $\mu$ (see [green] cross fora $\mu \approx 2.11$ in the case of $s=0.3$ considered in the
figure). This, in turn, produces an instability due to the
opposite
Krein sign of the colliding modes, yielding
a quartet of complex eigenfrequencies.
The (positive) imaginary part of the latter, is shown
in the bottom panel of Fig.~\ref{fig3}; see also the second
and third row of panels
for a typical profile and spectral plane of the relevant configuration.

The above features of the anomalous
mode seem structurally similar to the linear
periodic potential case, where again the cosinusoidal
case was found to be dynamically stable, while the sinusoidal one
was unstable beyond a critical lattice strength 
\cite{law,PGK:MPLB:04,vortex_JPB}.
These results also motivate an investigation of
how this phenomenology may be modified
in the presence of a dissipative term.

In this case, the pertinent model is the so-called dissipative GPE, which can be
expressed in the following dimensionless form:
\begin{eqnarray}
(i-\gamma) u_t=-\frac{1}{2} \Delta u + V(r) u + |u|^2 u - \mu u,
\label{veq14}
\end{eqnarray}
where the dimensionless parameter $\gamma$ can be
associated with the system's temperature in SI units according to \cite{gard2,gard}
(see also \cite{npprev})
\begin{equation}
\gamma = G\times \frac{4ma^2 k_{B}T}{\pi \hbar^2},
\label{gamma}
\end{equation}
with $k_B$ being Boltzmann's constant and the heuristically introduced
dimensionless prefactor $G \approx 3$. 
Note that in the dissipative model the interaction 
between the thermal cloud and the condensate is only modeled by
particle exchange resulting in the dissipative factor $\gamma$; we
should note that physically, the relevant case is that of $\gamma \ll
1$, although for illustration purposes we will occasionally show also
the results away from that regime. The chemical potential and trap
strength in Eq.~(\ref{veq14}) are set to the values $\mu=1$ and $\Omega=0.2$
(per the above discussion, it is understood how different $\mu$
and $\Omega$ will modify the relevant phenomenology).

\begin{figure}[tbp]
\begin{center}
\includegraphics[width=8cm,angle=0,clip]{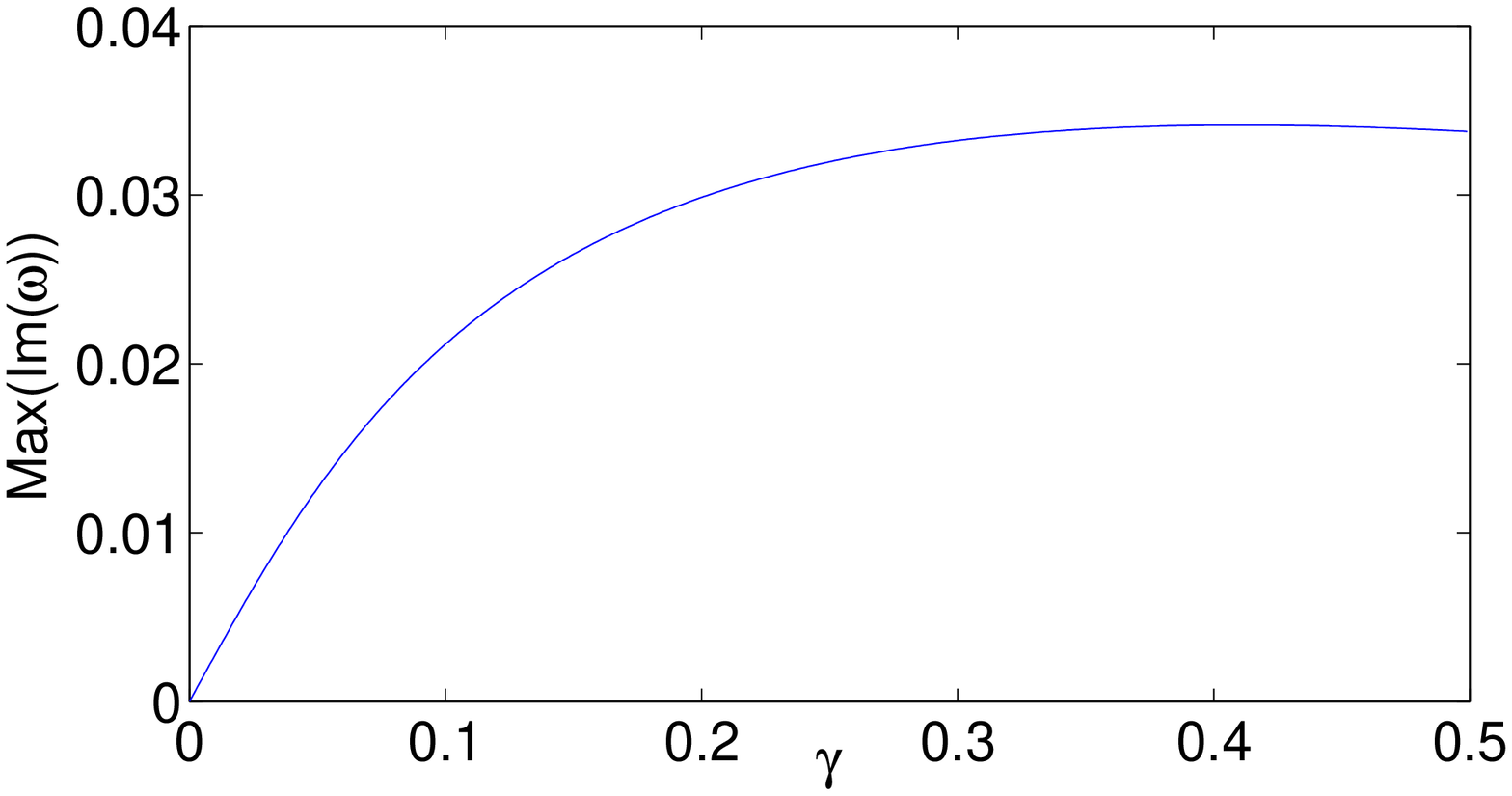}
\\
\includegraphics[width=8cm,angle=0,clip]{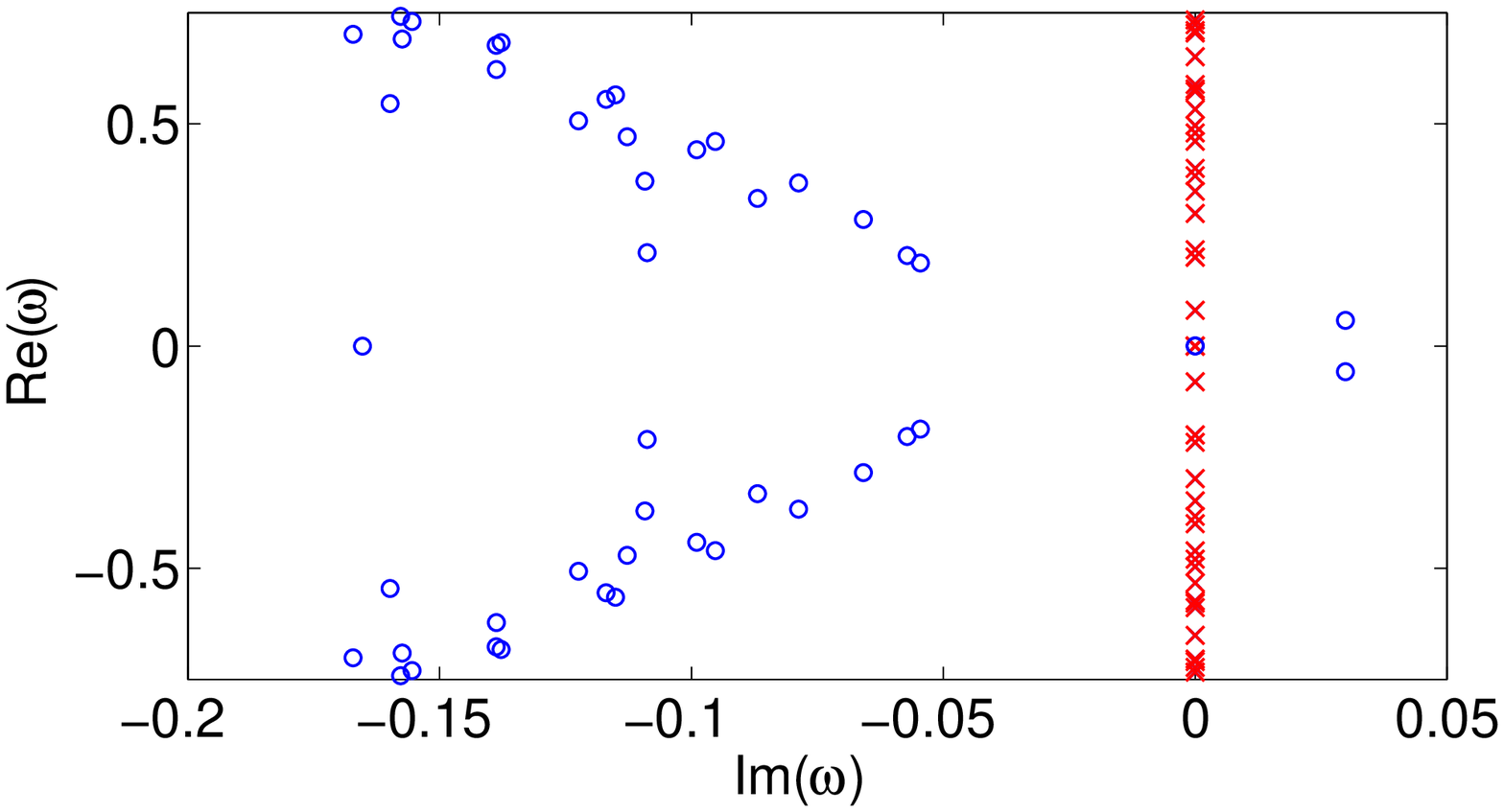}
\caption{(Color online) The top panel shows the immediate acquisition of a
non-vanishing imaginary part of the eigenfrequency associated with the anomalous
mode, as soon as the temperature-dependent
dissipative prefactor $\gamma$ becomes nonzero.
The bottom panel
highlights the special behavior of the anomalous mode by illustrating
the BdG spectrum for the cases of $\gamma=0$ (red crosses) and that
of $\gamma=0.2$ (blue circles). Notice how in the latter case the
eigenfrequencies form nearly two symmetric arcs in the negative
imaginary half-plane.}
\label{fig4}
\end{center}
\end{figure}

In Fig.~\ref{fig4}, we show the BdG spectrum of a vortex for
a case of $\gamma=0$ (zero temperature, i.e.,
no dissipation)
and for the case of $\gamma=0.2$ (finite temperature, i.e.,
dissipation).
It is clear that the lowest frequency mode of the condensate
(which for $\mu=1$ is the anomalous mode) is the one that, for
nonzero values of $\gamma$, immediately acquires a positive imaginary
eigenfrequency, contrary to what is the case for all other eigenmodes
of the system. In fact, precisely this property was rigorously
proved in Ref.~\cite{bjorn} for negative Krein sign eigenmodes, namely
that their bifurcation
upon such dissipative perturbations happens {\it oppositely} to that
of all other modes (of positive energy) of the system. This remarkable
feature is directly consonant with the property of this excited
state of the system resulting, via the effect of dissipation (and
through the complex nature of the relevant eigenmode) eventually
into the ground state of the system. Moreover, notice that this
complex eigenmode also implies the combination of growing amplitude
with the previously analyzed precessional motion, leading to the
spiraling of the vortex core toward the edges of the
TF cloud and its eventual disappearance, in favor of the ground state
of the system (see below).

\begin{figure}[tbp]
\begin{center}
\includegraphics[width=6cm,angle=0,clip]{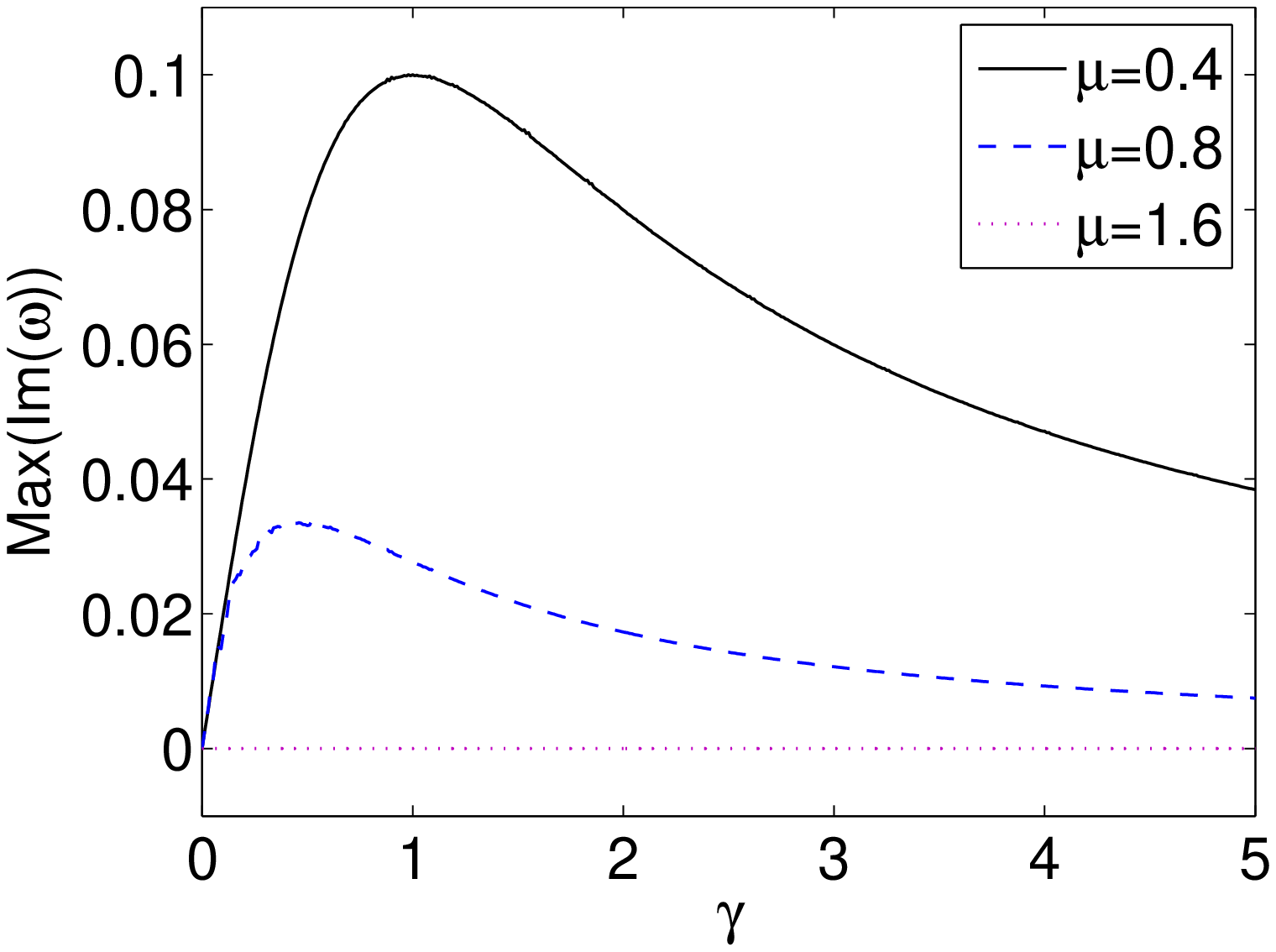}
\\[1.0ex]
~\includegraphics[width=5.8cm,angle=0,clip]{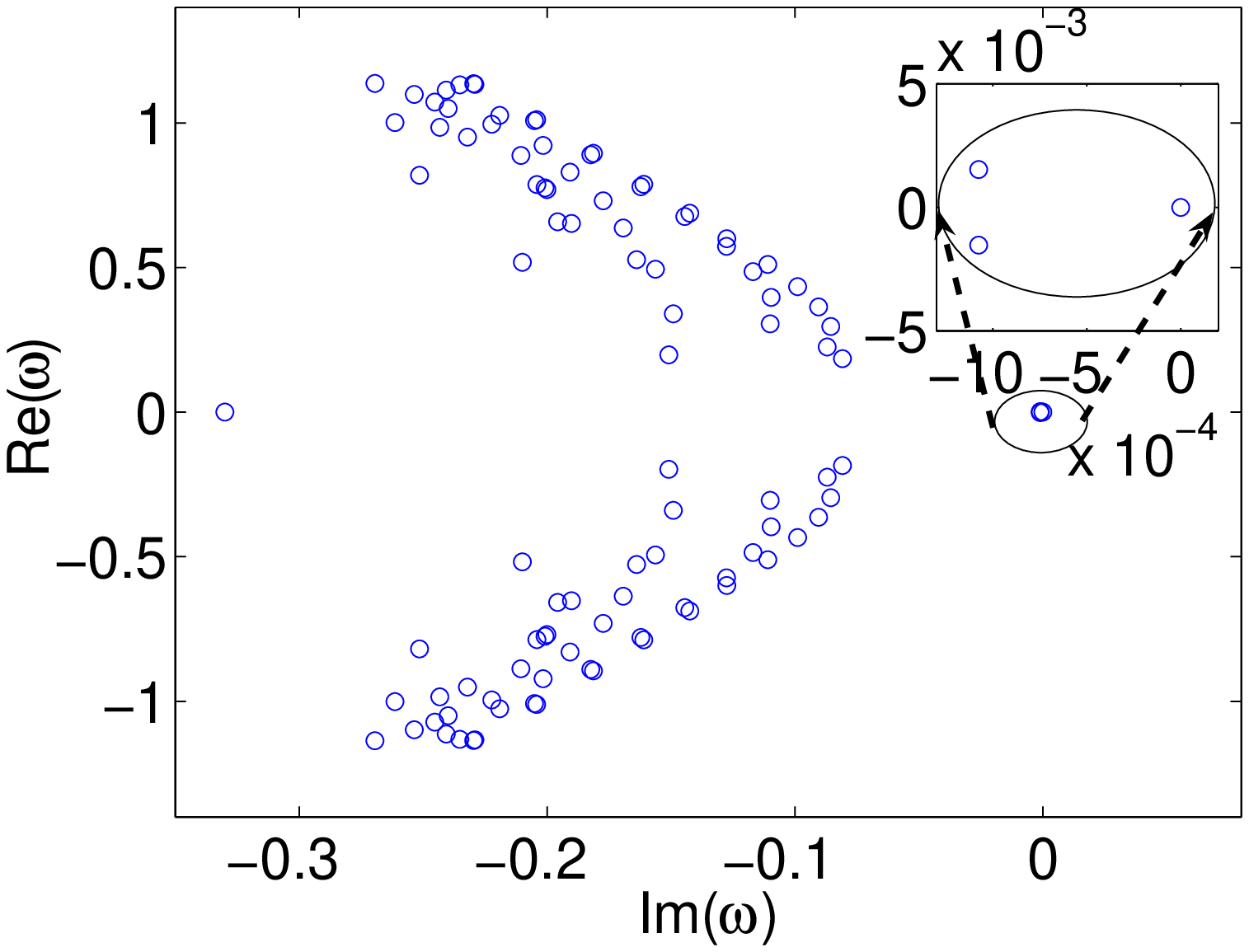}
\caption{(Color online) The top panel shows the immediate acquisition of a
non vanishing imaginary part of the eigenfrequency associated with the anomalous
mode, as soon as the temperature-dependent
dissipative prefactor $\gamma$ is nonzero for small chemical potentials,
but no acquisition of an imaginary part for large chemical potentials. The bottom
panel gives an overview of the eigenfrequencies for $\gamma=0.2$ and $\mu=1.6$.
The maximum imaginary part is zero. The inset shows that the eigenfrequencies
corresponding to the anomalous mode got shifted into the negative imaginary half-plane.}
\label{fig5}
\end{center}
\end{figure}

Lastly, let us
investigate the effect of a periodic modulation of the nonlinearity
on the stability of the system for the finite-temperature case.
For a periodic cosinusoidal modulation of the nonlinearity the
imaginary part of all eigenfrequencies vanishes and the system is stable
for $\gamma=0$ as discussed above.
The top panel of Fig.~\ref{fig5} shows the maximal
imaginary part of the eigenfrequency as a function of $\gamma$ for
different chemical potentials $\mu$ for $g(x,y)=1+ s \left(\cos^2(\pi x/4)
+ \cos^2(\pi y/4)\right)$, with $s=0.3$. For $\gamma=0$
the imaginary part of the eigenfrequency vanishes for all cases.
However, for small $\mu$ the imaginary part of the eigenfrequency
becomes non-zero immediately,
similar to the case of a constant nonlinearity strength. On the other hand,
for large chemical potential the maximal imaginary part of the
eigenfrequencies remains zero {\it independent} of $\gamma$.
Thus, the system remains
stable even in the presence of dissipation. This behavior can
be understood by investigation of Fig.~\ref{fig3}.
The occurrence of a positive imaginary part of the eigenfrequencies is due
to the fact that the anomalous mode is of negative Krein sign.
However, for the case of a cosinusoidal modulation of the nonlinearity
one observes that the value of the frequency of the anomalous mode decreases with
increasing $\mu$ and, finally, even crosses the origin. At that critical
point, the frequency curve shows a crossover with its opposite-value
companion (of the same pair). However the latter mode has a positive
Krein sign and therefore (since all negative signatures arise for
negative frequencies, and positive signatures for positive frequencies),
the imaginary parts of the eigenfrequencies become negative in the case of
nonzero $\gamma$.
The bottom panel in Fig.~\ref{fig5} provides an overview of the eigenfrequencies for
$\gamma=0.2$ and $\mu=1.6$. All eigenfrequencies were shifted further
into the negative imaginary half-plane in comparison to the corresponding panel
in Fig.~\ref{fig4}. Importantly, the eigenfrequencies corresponding to the
(formerly) anomalous mode got shifted into the negative imaginary half-plane as
is shown in the inset.

\subsection{Direct Numerical Simulations}
In this section we show results obtained by direct numerical
integration of Eq.~(\ref{veq1}) starting with different initial states
containing a single vortex.
In order to determine the position of the vortex as a function of time
we first compute the fluid velocity \cite{Jackson98}
\begin{equation}
\mathbf{v}_s=-\frac{i}{2} \frac{u^\star \mathbf \nabla u - u \mathbf \nabla u^\star}{|u|^2}.
\end{equation}
The fluid vorticity is then defined as
$\mathbf{\omega}_{\rm vor}=\mathbf \nabla \times \mathbf{v}_s$. Due
to our setup,
the direction of the fluid vorticity is always the
$z$-direction and, therefore, we can treat this quantity as a scalar.
Furthermore, we investigate single vortex states leading to a single maximum of the fluid vorticity at the position of the vortex. This allows us to determine the position of the vortex by determining
the center of mass of the vorticity $\mathbf{\omega}_{\rm vor}$.

\begin{figure}[tbp]
\begin{center}
\includegraphics[width=8cm,angle=0,clip]{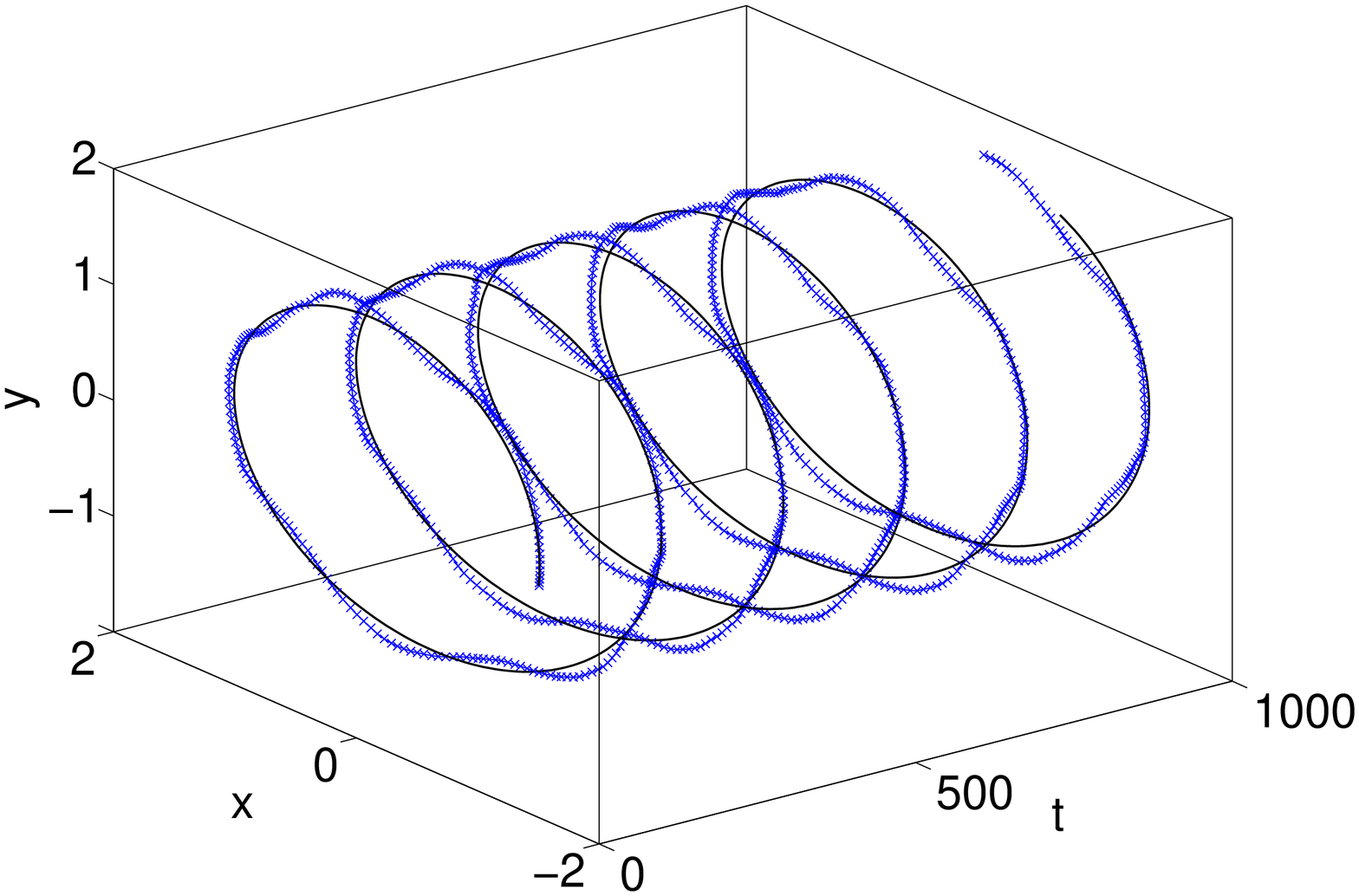}
\includegraphics[width=8cm,angle=0,clip]{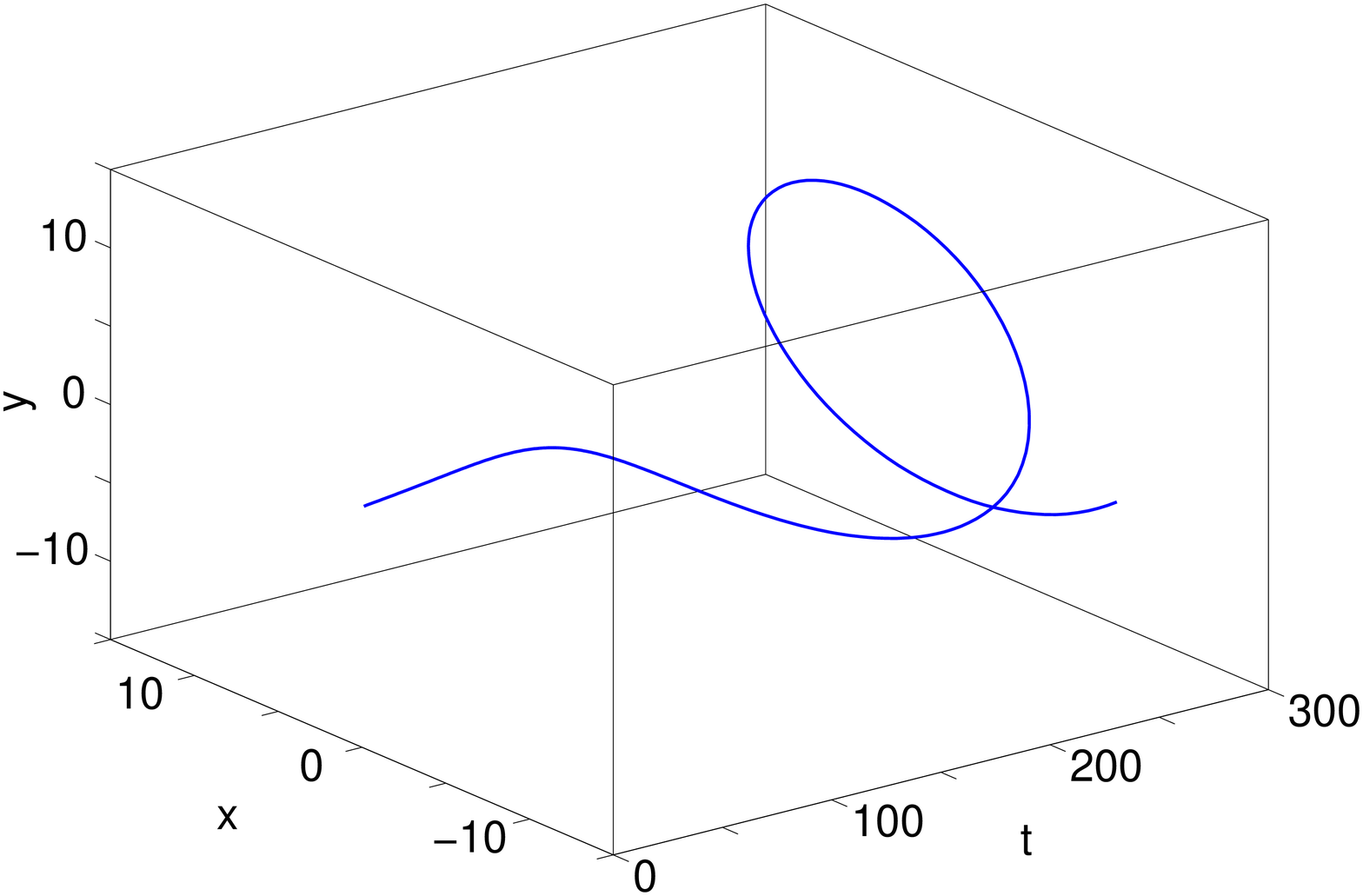}
\caption{(Color online) 
The top panel shows the trajectory of the vortex for $g=1$
and $\mu=3$ obtained by direct integration (crosses) and the theoretical
prediction (solid line) obtained by solving Eqs.~(\ref{veq8})--(\ref{veq9}).
Notice the excellent agreement between the two.
The bottom panel shows the corresponding trajectory for the case taking
into account dissipation, namely integrating Eq.~(\ref{veq14}) with $\gamma=0.2$.}
\label{fig6}
\end{center}
\end{figure}

Figure \ref{fig6} shows the evolution of a single vortex for $g=1$. We
displaced the vortex initially from the center of the trap to
$(x_0,y_0)=(-1.5,0)$ and propagated the state numerically using
Eq.~(\ref{veq1}). The thus obtained results are compared to the
solutions of Eqs.~(\ref{veq8})--(\ref{veq9}) $x=x_0 \cos(C t)$ and
$y=y_0 \sin(C t)$ with $C=({\Omega^2}/{(2 \mu)}) \log(A
{\mu}/{\Omega})$ and the initial position $(x_0,y_0)$. The
theoretical predictions agree very well with our numerical
findings: the vortex oscillates around the center of the trap with
constant frequency and radius (see top panel).
%
The bottom panel shows the trajectory for the case of constant
nonlinearity $g=1$ but for finite temperature (i.e., including dissipation).
The results shown were obtained by direct numerical integration of
Eq.~(\ref{veq14}) for $\gamma=0.2$, with the initial condition
being a slightly perturbed eigenstate of the system. Due to the instability of the
system this small perturbation leads to the spiraling out of the vortex,
as is physically anticipated in the presence of finite temperature
\cite{prouk2}; we note in passing that this work
contains a detailed model from microscopic
first principles that illustrates a similar phenomenology upon
a spatially dependent inclusion of the coupling of
the condensate with the thermal cloud.

\begin{figure}[tbp]
\begin{center}
\includegraphics[width=6cm,angle=0,clip]{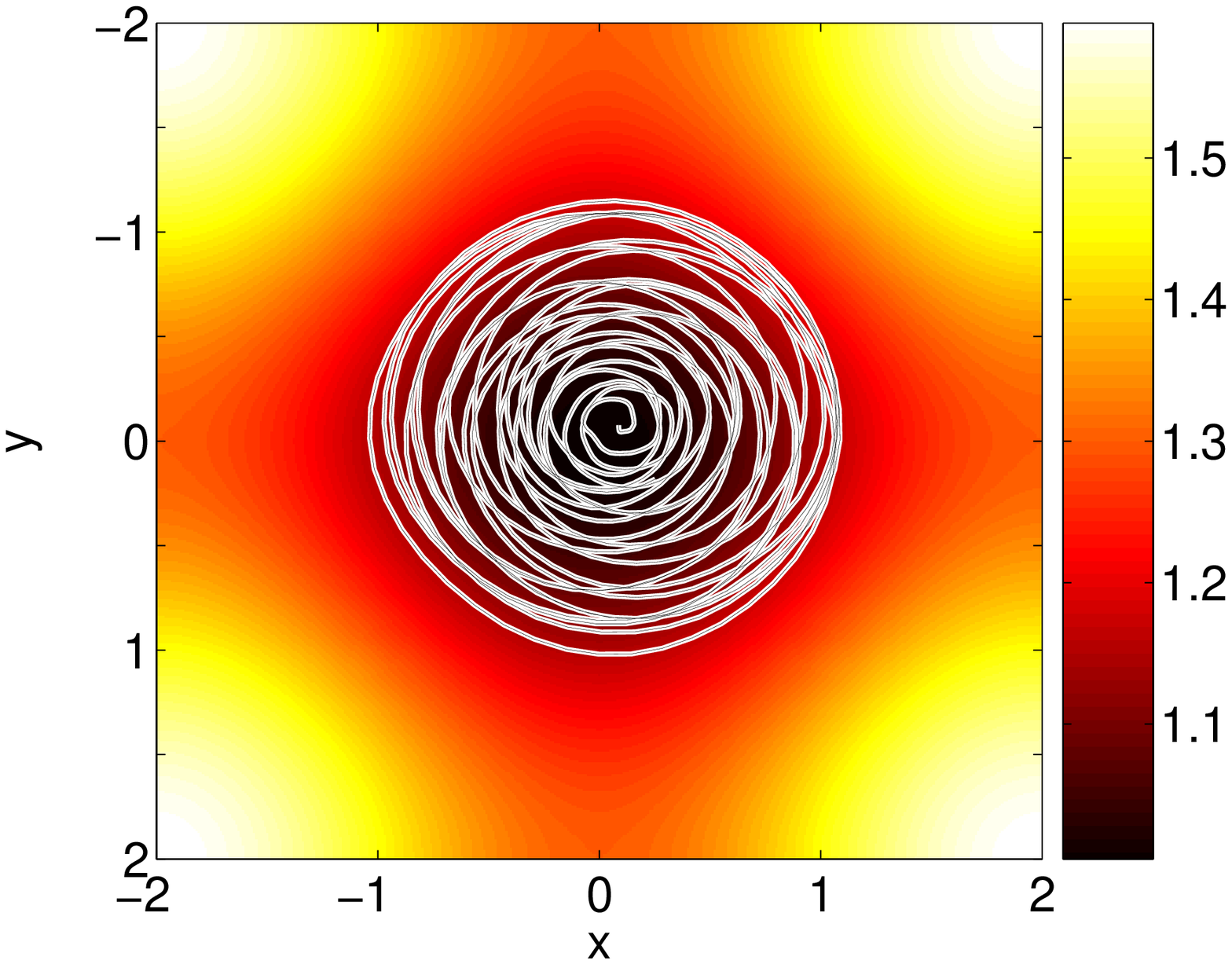}
\\
\includegraphics[width=6cm,angle=0,clip]{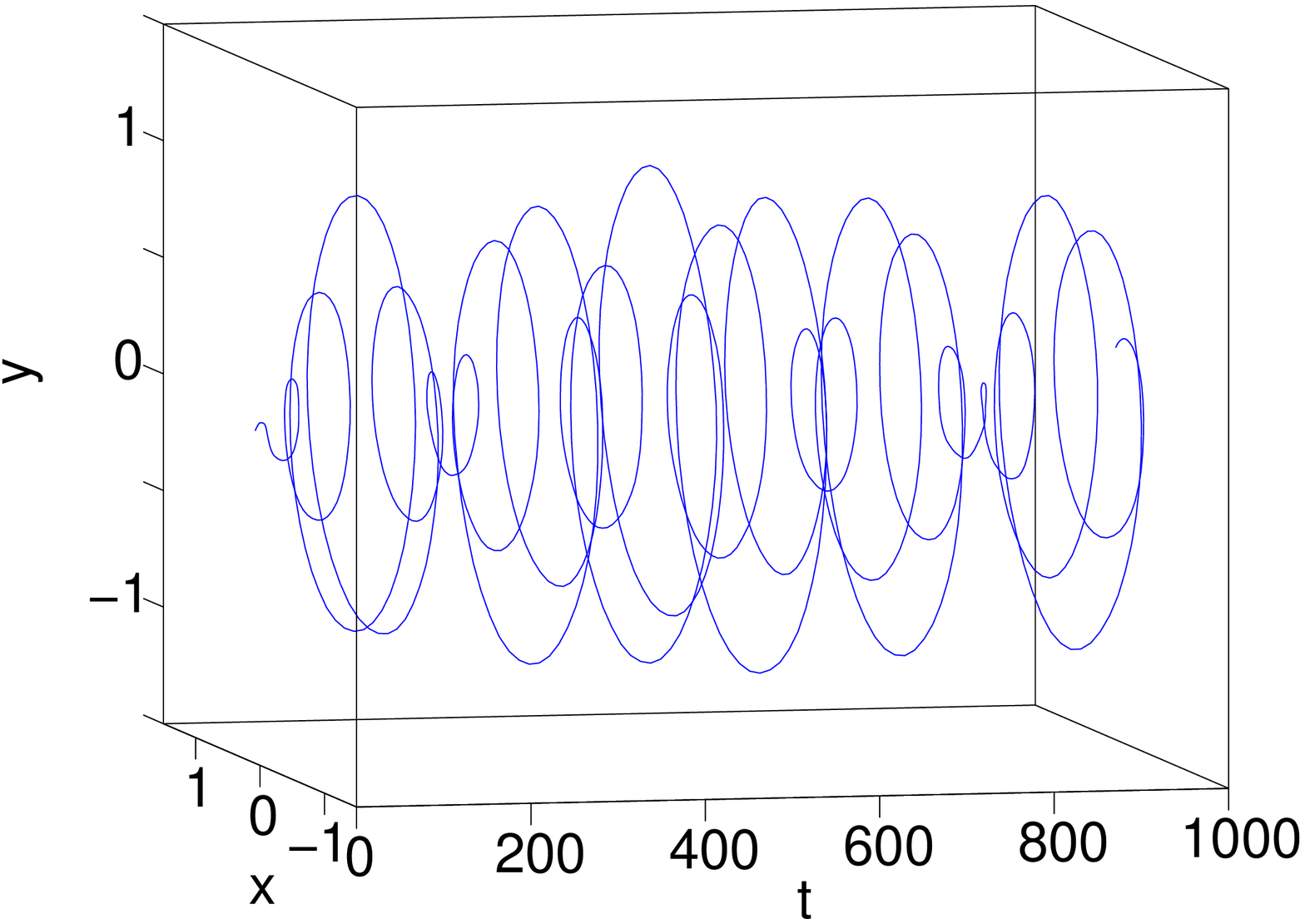}
\caption{(Color online) The trajectory of a vortex for $\mu=3$ for the case
$g(x,y)=1+ s \left(\sin^2(\pi x/4)+ \sin^2(\pi y/4)\right)$ with
$s=0.3$. In the top panel the trajectory is plotted on top of the
profile of the coupling $g(x,y)$,
whereas the bottom panel shows the trajectory as a function of time.}
\label{fig7}
\end{center}
\end{figure}

Figure \ref{fig7} shows the trajectory of a vortex for the case of a
periodically modulated sinusoidal nonlinearity, $g(x,y)=1+ s \left(\sin^2(\pi
x/4)+ \sin^2(\pi y/4)\right)$. The initial configuration is a slightly
perturbed eigenstate leading to a small shift of the position of the
vortex. Due to the instability of the sinusoidal $g(x,y)$
landscape, the vortex spirals outwards initially, but then spirals inwards
after reaching a region with approximately constant nonlinearity.
Subsequently the vortex follows a series of such alternating
(spiraling first outwards, and then inwards) cycles in an
apparently quasi-periodic orbit.

\begin{figure}[tbp]
\begin{center}
\includegraphics[width=4.2cm,angle=0,clip]{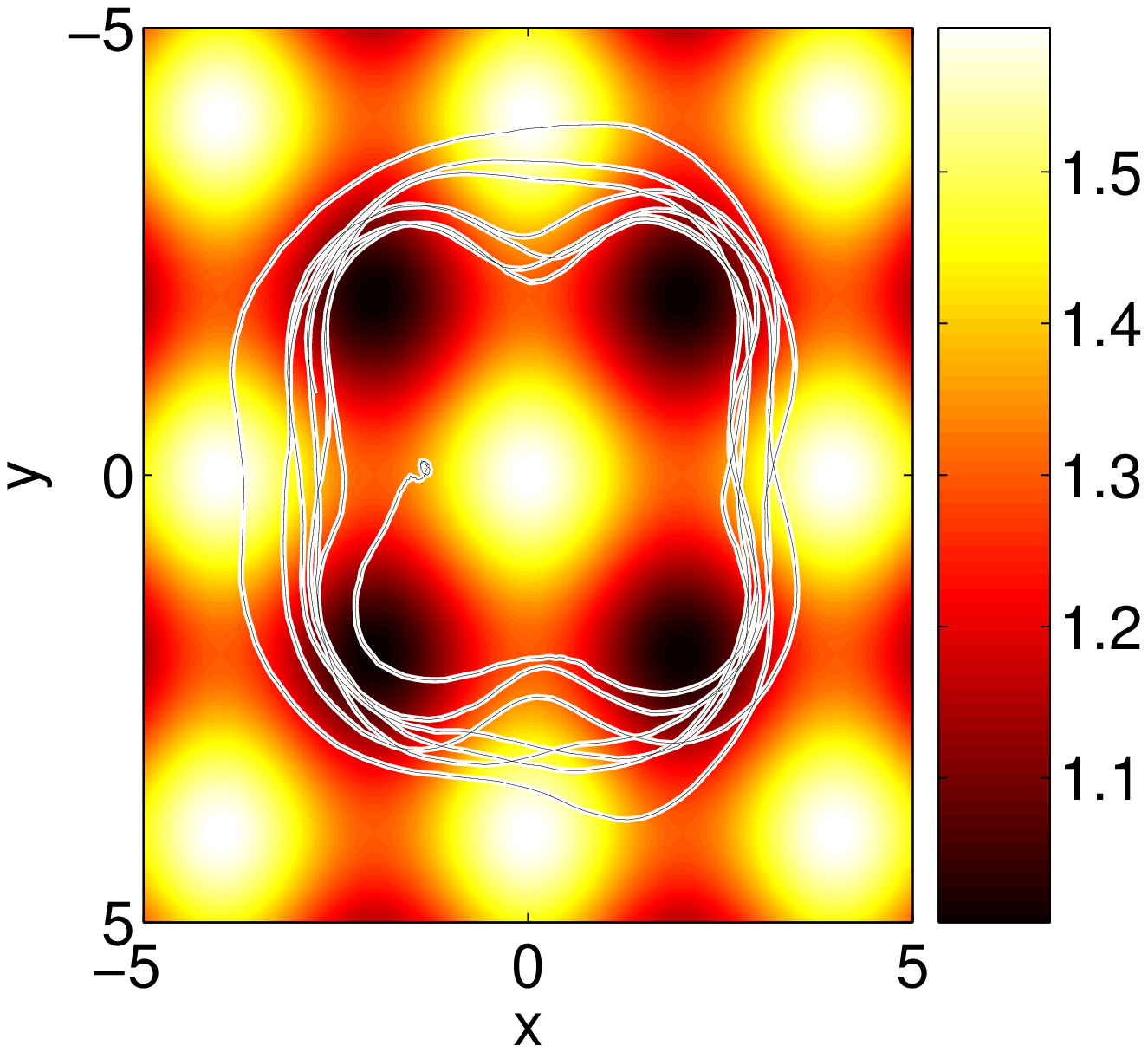}
\includegraphics[width=4.2cm,angle=0,clip]{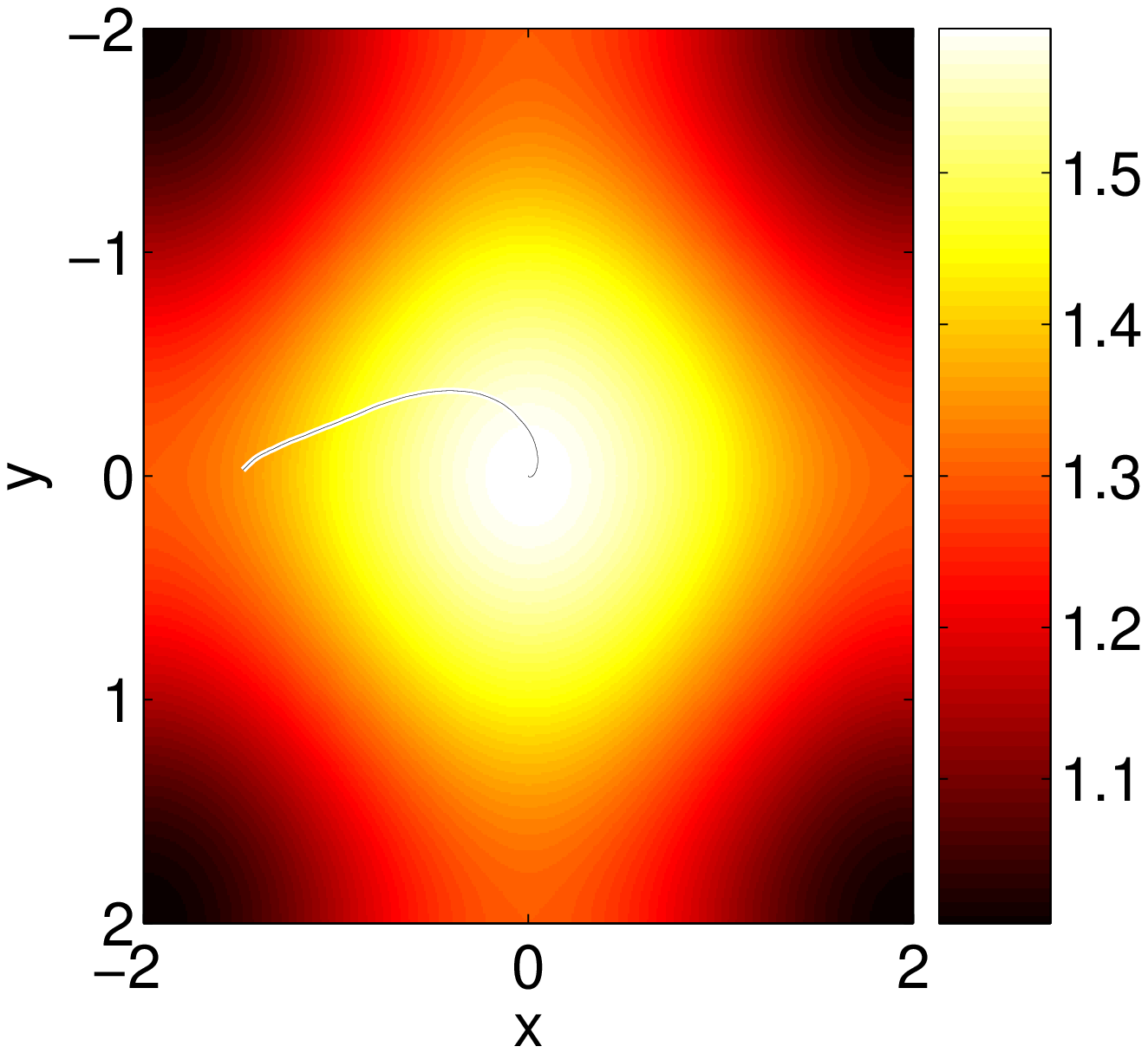}
\caption{(Color online) The trajectory of the vortex for $\mu=3$ for the case
$g(x,y)=1+ s \left(\cos^2(\pi x/4)+ \cos^2(\pi y/4)\right)$
with $s=0.3$
without dissipation (left) and with dissipation (right) with
$\gamma=0.2$. The trajectories are plotted on top of the profile of
the coupling $g(x,y)$. Notice the fast inward vortex motion in the
presence of dissipation.}
\label{fig8}
\end{center}
\end{figure}

Figure \ref{fig8} shows the trajectory of a vortex for the case of a
periodically modulated cosinusoidal
nonlinearity, $g(x,y)=1+ s \left(\cos^2(\pi x/4)+ \cos^2(\pi
y/4)\right)$, without dissipation (left panel) and with dissipation (right panel).
In this case, small perturbations do not get amplified since the
system is stable. However, a macroscopic displacement of the
vortex to $(x_0,y_0)=(-1.5,0)$ leads to the
trajectories shown in the figure (see left panel).
In the case without dissipation the vortex moves outwards
(reaching a region outside the ``square'' of the first minima of
the nonlinearity) and oscillates around the center on a trajectory with
roughly constant nonlinearity. For the case with cosinusoidal
nonlinearity {\it and} dissipation (see right panel),
the vortex remains stable against small perturbations and does not
spiral out, contrary to the case of a constant nonlinearity.
Even for a macroscopical displacement the vortex moves back to the
center of the trap and remains stable there.
This behavior is possible because the effective potential due
to the spatial variation of the nonlinearity creates the possibility
for a metastable vortex state {\it even in the presence of dissipation}.

In conclusion, the modulation of the nonlinearity opens up the
possibility to stabilize the vortex against excitations due to finite
temperature effects. This can be extremely useful for setups which require
stable vortex states for a long period of time as, e.g., in the recent
work of Ref.~\cite{thanvanthri} which suggests
the use of a superposition of two counter rotating BECs as a gyroscope.

\section{Conclusions}
\label{secIV}

In summary, in the present work, we examined the role of anomalous
modes in the motion of vortices in harmonically confined condensates.
We have also focused on the settings of spatially dependent scattering
lengths and of finite temperature (as well as the combination thereof).
We found a number of interesting results,
including an explicit semi-analytical expression for the precession
frequency in the trap (by means of the matched asymptotics technique),
which was found to be in excellent agreement with both bifurcation
and direct numerical integration results, for different chemical
potentials and trap frequencies within the Thomas-Fermi regime.

We subsequently examined how the spectrum (more generally ---and
the anomalous mode in particular) are affected
by the presence of spatially-dependent (harmonic) interatomic interactions.
We found that the latter may induce or avoid instabilities depending on the curvature
and the strength of the nonlinearity variation. The effect of
temperature was examined in a simple phenomenological setting which,
however, still enabled us to observe the thermal instability
of the vortex and its rapid spiraling towards the edges of
the condensate cloud.
Intriguingly enough, we also demonstrated that
the effect of spatially dependent nonlinearities may avoid
the thermal instability of the vortex by creating a local
metastable effective energy minimum wherein the vortex can spiral
inwards towards the center of the harmonic trap.

It would be particularly interesting to try to extend both the
analytical and the numerical considerations herein towards
different directions. On the one hand, it is appealing
to find similar particle-like equations for the motion
(and interaction) of multiple vortices within
the parabolic trap. On the other hand, it would be especially
relevant to consider such multi-core realizations in the
presence of the thermal and spatially dependent nonlinear
effects. Yet another direction could be to extend considerations
presented herein to the case of vortex rings and
their dynamics.
Such studies are presently in progress.

\section*{Acknowledgments}

P.G.K.~gratefully acknowledges support from
the NSF-CAREER program (NSF-DMS-0349023), from NSF-DMS-0806762
and from the Alexander von Humboldt Foundation. 
R.C.G.~gratefully acknowledges
the hospitality of the Grupo de F\'{\i}sica No Lineal (GFNL) of
Universidad de Sevilla and
support from NSF-DMS-0806762,
Plan Propio de la University de Sevilla and
Grant\#IAC09-I-4669 of Junta de Andaluc\'{\i}a.
P.G.K. and S.M.  acknowledge a number of useful discussions with Kody Law.
The work of D.J.F.~was partially supported by the Special Account for 
Research Grants of the University of Athens.

\section*{Appendix: Equations of Motion for the Vortex}

In this Appendix we detail the effects of the potential on the
position of the vortex via a matched asymptotics approach
between an inner and an outer perturbative solution.
The inner solution is of the form:
\begin{eqnarray}
u(r,\theta)=\left[u_0(r) + \varepsilon \chi(r) \cos(\theta) \right]
e^{i \left[S \theta + \varepsilon \eta(r) \sin(\theta) \right]},
\label{veq2}
\end{eqnarray}
where $\varepsilon$ is a formal small parameter associated with
the slow speed of precession and $u_0(r)$ is the radial vortex profile,
while $\chi$ and $\eta$ are functions
of $r$, whose asymptotics have been detailed in Refs.~\cite{pis1,pis2}
(see also Ref.~\cite{fetter1}).
The outer perturbative solution can be
obtained by a lowest-order equation for the phase, resulting
from a rescaling of
space, $r \rightarrow \varepsilon r$, and
time, $t \rightarrow \varepsilon^2 t$, namely:
\begin{eqnarray}
\Delta \theta + {\bf F} \cdot \nabla \theta=0,
\label{veq3}
\end{eqnarray}
where ${\bf F}=\nabla \log(|u_b|^2)$
(hereafter, boldface is used to denote vectors) and
$|u_b|^2$ is the (background, hence the relevant subscript)
BEC density in the absence of the vortex;
notice that the density can be approximated in the Thomas-Fermi (TF) limit
as $|u_b|^2=\mu-V(r)$.

Interestingly, the similarity of Eq.~(\ref{veq3})
to Eq.~(20) of Ref.~\cite{pis2} could lead
to the impression that the detailed formalism
of Ref.~\cite{pis2} could be blindly followed giving
rise to the precessional motion of Eq.~(27) therein.
However, this turns out to be incorrect. Particularly, in Ref.~\cite{pis2},
it was non-generically assumed that ${\bf F}$ of the outer
expansion
can be accurately approximated by a constant.
In our case where ${\bf F} \approx -\Omega^2 {\bf r}/\mu$
(for small and intermediate distances where the matching
with the inner expansion is performed), this approximation is
clearly not an appropriate one. Instead, we follow the original
formulation of Ref.~\cite{pis1}, which employs the change of variables
$\phi(x,y) \rightarrow \theta(x,y)$:
\begin{eqnarray}
\theta_x =-S \left(\phi_y- \phi \frac{\Omega^2}{\mu} y \right),
\label{veq4}
\\
\theta_y=\phantom{-}S \left(\phi_x-\phi \frac{\Omega^2}{\mu} x \right),
\label{veq5}
\end{eqnarray}
(we will suppress the $S$-dependence hereafter, focusing on
singly-charged vortices). Then, the equation for $\phi$ reads:
\begin{eqnarray}
\Delta \phi-\frac{\Omega^2}{\mu} (x \phi_x + y \phi_y) -
2 \frac{\Omega^2}{\mu} \phi = 0,
\label{veq6}
\end{eqnarray}
which, upon using the transformation $\phi=H(r)/\sqrt{\mu-V(r)}$, yields
\begin{eqnarray}
\Delta H - \frac{\Omega^2}{\mu} H=2 \pi \delta({\bf r}-{\bf r}_0),
\label{veq7}
\end{eqnarray}
assuming a point vortex source at ${\bf r}_0$. This leads to the
asymptotic behavior $H=-K_0(m |{\bf r}-{\bf r}_0|)$, where $K_0$
is the modified Bessel function and $m=\Omega/\sqrt{\mu}$.
This should be directly compared
with Eq.~(23) of Ref.~\cite{pis2}, showcasing that instead of $F^r/2$
in the latter equation, here we have the constant factor $m$ multiplying
the distance from the vortex core. Once this {\it critical}
modification is made, the rest of the calculation of Ref.~\cite{pis2} can be
followed directly, yielding the final result (in the presence of
the trap):
\begin{eqnarray}
\dot{x}_v&=&\phantom{-}\frac{\Omega^2}{2 \mu} \log \left(A \frac{\mu}{\Omega}\right) y_v,
\\
\dot{y}_v&=&-\frac{\Omega^2}{2 \mu} \log \left(A \frac{\mu}{\Omega}\right) x_v,
\end{eqnarray}
where the pair $(x_v,y_v)$ defines the location of the vortex center,
$A$ is an appropriate numerical factor (detailed comparison
with numerics yields very good agreement in the TF regime e.g.~for
$A \approx 8.88\approx 2 \sqrt{2}\pi$, see Sec.~II).

It should be noted that this equation is valid for
  small displacements from the trap center which lends further 
support to the connection of this dynamics with the relevant mode
of the BdG analysis. For larger displacements from the vortex center,
this expression should be appropriately corrected \cite{McGee}.


\end{document}